\documentclass[%
 reprint,
superscriptaddress,
 amsmath,amssymb,
 aps,
pra,
]{revtex4-2}

\usepackage{graphicx}
\usepackage{dcolumn}
\usepackage{bm}
\usepackage{tabularx}
\usepackage{braket}
\usepackage{xtab,afterpage}
\usepackage{multirow}
\usepackage{booktabs}
\usepackage{xcolor}

\usepackage{hyperref}
\hypersetup{colorlinks=true, citecolor=blue, urlcolor=blue, linkcolor=blue}

\def\ypo4{YPO$_4$}
\def\d#1/d#2{\frac{\partial #1}{\partial #2}}

\def\basisu#1s#2p#3d{\hbox{\it (#1s#2p#3d)\ }}
\def\basis#1s#2p#3d/#4s#5p#6d{\hbox{\it (#1s#2p#3d)/[#4s#5p#6d]\ }}

\usepackage{soul}

\def\ypo4{YPO$_4$}
\newcounter{tablfootnote}
\def\fn#1{\setcounter{tablfootnote}{#1}%
          \raise0.5ex\hbox{\ \footnotesize{\alph{tablfootnote}%
}}}

\makeatletter
\def\yo8{YO$_8$@CTEP$_{\rm min}$}
\def\ypo46{Y(PO$_4$)$_6$@CTEP$_{\rm ext}$}
\makeatother
\newcounter{enumsli}

\begin{document}

\title{Finite-order method to calculate approximate density matrices\\ in the Fock-space multireference coupled cluster theory}

\author{Alexander~V.~Oleynichenko}
\email{oleynichenko\_av@pnpi.nrcki.ru}

\affiliation{B. P. Konstantinov Petersburg Nuclear Physics Institute of National Research Center ``Kurchatov Institute'' (NRC ``Kurchatov Institute'' -- PNPI), Gatchina, Leningrad district 188300, Russia}

\affiliation{Moscow Institute of Physics and Technologies (National Research University), Institutskij pereulok 9, Dolgoprudny, Moscow region 141700, Russia}

\author{Andr\'ei~Zaitsevskii}
\email{zaitsevskii\_av@pnpi.nrcki.ru}

\affiliation{B. P. Konstantinov Petersburg Nuclear Physics Institute of National Research Center ``Kurchatov Institute'' (NRC ``Kurchatov Institute'' -- PNPI), Gatchina, Leningrad district 188300, Russia}

\affiliation{Department of Chemistry, M.~V. Lomonosov Moscow State University, 119991 Moscow, Russia}

\author{Leonid~V.~Skripnikov}
\email{skripnikov\_lv@pnpi.nrcki.ru}

\affiliation{B. P. Konstantinov Petersburg Nuclear Physics Institute of National Research Center ``Kurchatov Institute'' (NRC ``Kurchatov Institute'' -- PNPI), Gatchina, Leningrad district 188300, Russia}

\affiliation{Saint Petersburg State University, 7/9 Universitetskaya nab., 199034 St. Petersburg, Russia}

\author{Ephraim~Eliav}
\email{ephraim@tau.ac.il}

\affiliation{School of Chemistry, Tel Aviv University, Tel Aviv 6997801, Israel}

\date{\today}

\begin{abstract}
An efficient approach to calculate approximate pure-state and transition reduced density matrices in the framework of the multireference relativistic Fock-space coupled cluster (FS CC) theory is proposed. The method is based on the effective operator formalism and consists of the direct substitution of the FS CC Ansatz for a wave operator into the effective operator expression with the subsequent truncation of expansion at the terms quadratic in cluster amplitudes. The final density matrix is defined by active-space density matrices of different ranks ``dressed'' with contributions from cluster operators. The method gives a connected expression for pure-state density matrices, provided that the intermediate normalization condition is fulfilled. Moreover, under some additional assumptions, the connectivity can also be ensured for calculated transition property matrix elements and natural transition spinors. The developed technique allows for fast and accurate calculations of one-particle reduced density matrices for a wide range of electronic states. A pilot application of the new technique to construct averaged atomic natural orbital (ANO) basis sets for fully relativistic electronic structure calculations is presented.
\end{abstract}

\maketitle

\section{Introduction}

Evaluation of reduced density matrices (RDMs) is a long-standing problem in electronic structure theory~\cite{Davidson:76}. An electronic structure method can be particularly useful for a wide range of practical problems beyond single-point energy calculations, provided that an algorithm for constructing density matrices is developed. Probably one of the most important and well-known applications of RDMs is the calculation of energy derivatives with respect to positions of atomic nuclei; this procedure underlies an optimization of molecular geometry and requires one- and two-particle reduced density matrices (further denoted as 1-RDM and 2-RDM, respectively)~\cite{Koch:90,Helgaker:00,Gauss:Lambda:91,ShavittBartlett:09}. Secondly, RDMs can be useful for evaluating matrix elements of property operators. Transition RDMs are used to calculate off-diagonal matrix elements defining electronic transition probabilities, e.~g. transition electric or magnetic dipole moments. Furthermore, density matrices are of interest in performing analysis of electronic distributions and bonding in molecules and solids. For example, the diagonalization of a pure-state 1-RDM yields natural orbitals (or spinors in the relativistic case), which can be further visualized or used in some other way to analyze electronic structure and chemical bonding in a given electronic state (see~\cite{Krylov:NTO:20} and references therein). Transition density matrices can be treated similarly, resulting in a compact and clear representation of electronic transition in terms of natural transition orbitals (spinors)~\cite{Martin:NTO:03,Ivanov:19,Krylov:NTO:20,Isaev:24}. Natural spinors obtained within a rather simple approximation can serve as a reduced-size one-electron basis for higher-level correlated calculations~\cite{Jensen:MP2NO:88,Haldar:21,Chamoli:24}. One can also perform a local analysis of a one-particle density matrix to define effective atomic states in a compound (\cite{Titov:14} and references therein). Last but not least, RDMs obtained for several electronic states can be averaged to obtain flexible contracted basis sets for molecular calculations by the averaged atomic natural orbital (ANO) technique~\cite{Widmark:90,Almlof:91,Skripnikov:E120:13}.

While the construction of reduced density matrices is relatively straightforward for variational methods of electronic structure theory operating with explicitly given wavefunctions like configuration interaction, it becomes challenging for non-variational approaches and/or methods providing only a recipe for calculating wavefunctions rather than wavefunctions themselves~\cite{Helgaker:12,Norman:18}. Various formulations of the coupled cluster (CC) theory designed for electronic structure problems of either single-reference or multireference (MR) character fall into this category. The general approach to derive a method to calculate analytic density matrices for a given wavefunction Ansatz consists of constructing the Lagrangian function with its subsequent minimization with respect to some set of parameters defining an electronic wavefunction~\cite{Jorgensen:88,Helgaker:00,Helgaker:12}. This idea turned out to be very fruitful. To the moment nearly all versions of the coupled cluster theory are provided with algorithms for evaluating pure-state density matrices~\cite{Gauss:Lambda:91,Stanton:93,Szalay:95,Kallay:03,Kallay:Hessian:04,Kallay:04,Bartlett:07,Pittner:BW:gradients:07,Prochnow:09,Shee:16,Samanta:icMRCC:gradients:18}. It is worth mentioning the series of papers of Pal and co-workers, who proposed a route towards analytic pure-state one-particle density matrices for the Fock space multireference coupled cluster (FS CC) method (see~\cite{Shamasundar:04,Gupta:13} and references therein). The situation is different for transition density matrices. For the most widely used CC-based approaches aimed at the simulation of excited electronic states and their properties like the equation-of-motion (EOM) CC and linear-response (LR) CC, the analytic expressions for transition density matrices are well established~\cite{Stanton:93,Kallay:04,Wang:EOM:15,Halbert:21}. As for the genuine multistate MR CC approaches~\cite{Lyakh:11}, the most notable progress was achieved for the Fock-space coupled cluster method, for which an attempt to formulate a recipe to calculate transition dipole moments based on the minimization of the FS CC Lagrangian was reported~\cite{Bhattacharya:13,Bhattacharya:14}.

Among all versions of the coupled cluster theory oriented at the modeling of electronic states with pronounced multireference character, the Fock-space formulation seems to be one of the most well-established and widely used~\cite{Lindgren:87,Kaldor:91,Visscher:01,Musial:04,Eliav:Review:22}. Its relativistic production-level program implementations~\cite{Eliav:94,Saue:20,Oleynichenko:EXPT:20} served as tools to solve many problems in modern atomic and molecular physics. Latest applications of FS CC include first (and experimentally confirmed) predictions of spectra of transuranium and superheavy elements~\cite{Raeder:Nobelium:18}, spectroscopy of short-lived radioactive molecules~\cite{Zaitsevskii:RaF:22,Skripnikov:AcF:23,ArrowsmithKron:24}, studies of laser-coolable molecules~\cite{Isaev:RaOH:17,Osika:RaF:22,Oleynichenko:AcOH:22,Isaev:24,AthanasakisRaFPinning:2023}, extraction of nuclear electromagnetic moments~\cite{Skripnikov:Bi:21} and isotope shift parameters~\cite{Konig:Si:24} and even successful predictions of localized excitations in solids~\cite{Gomes:08,Oleynichenko:YPO4:24}. The success of the FS CC method can be partly explained by the relative simplicity of its working equations resulting in the possibility of efficient program implementation, and the existence of the intermediate Hamiltonian formulations aimed at solving the so-called intruder state problem inherent for MR CC with (quasi-)complete model spaces~\cite{Meissner:IH:98,Eliav:XIH:05,Musial:IH:08,Zaitsevskii:QED:22}. Moreover, the accuracy of FS CC can be systematically improved by including higher excitations (e.~g. the FS CCSDT model)~\cite{Musial:04,Dutta:15,Oleynichenko:CCSDT:20}. Accounting for higher excitations is essential for accurate simulations of electronic states with more than two unpaired electrons~\cite{Hughes:92,Eliav:Review:15,Skripnikov:Bi:21}. Finally, the list of tools available within the FS CC framework also includes two approximate approaches designed to calculate off-diagonal matrix elements of property operators, namely the finite-field~\cite{Zaitsevskii:Optics:18} and finite-order~\cite{Zaitsevskii:ThO:23,Oleynichenko:Optics:23} techniques.

The growing popularity and scope of applicability of the FS CC method clearly appeal to the search for efficient approaches to construct pure-state and transition density matrices. Thus, in the present work, we focus on a new economical method to calculate approximate density matrices in the FS CC theory. As mentioned above, a rigorous method to construct analytic FS CC density matrices for pure electronic states is, in principle, known~\cite{Shamasundar:04,Gupta:13}. For transition properties the bivariational approach to calculate transition dipole moments (but not transition density matrices) was proposed. All these developments implied non-relativistic approximations for the electronic Hamiltonian and were focused on the Fock space sectors with one particle over the closed-shell vacuum (the $0h1p$ sector), one hole ($1h0p$) or one hole and one particle simultaneously ($1h1p$). Typically, electronic states with such a determinantal structure can be reached within other well-developed approaches like EOM-CC, which readily provide analytic density matrices. These papers did not consider more versatile Fock space sectors with two or more particles over the closed-shell vacuum. Furthermore, within the FS CC method the so-called $\Lambda$-equations defining Lagrange multipliers are state-specific and must be solved separately for each electronic state under consideration. This circumstance is not an obstacle if only one or several electronic states are to be thoroughly studied, but one is interested in a broad manifold of electronic states for systems with two or more unpaired electrons, the problem can become unmanageable (for example, lanthanide ions typically have very dense electronic spectra including hundreds and thousands of excited states in the optical range).

For many practical applications, it is important to generate one-particle density matrices quickly and in a computationally efficient manner, even at the cost of a slight loss of accuracy. For example, such applied problems as calculations of natural (transition) orbitals or spinors, visualization issues, population analysis, construction of contracted basis sets, or fast estimates of property operator matrix elements do not require the use of exact density matrices. At the same time, it is more important to obtain them without significant additional computational expenses. In the present paper, we propose an approximate but still quite accurate approach to calculating density matrices within the FS CC framework. It is based on directly substituting an exponential FS CC Ansatz for a wave operator into expressions defining density matrices with the subsequent truncation of the resulting series. Thus, the proposed method complements the finite-order technique derived recently to calculate expectation values and transition moments of one-electron property operators~\cite{Oleynichenko:Optics:23}; the $\Lambda$-equations are completely avoided in this approximate scheme. It is worth noting that the developed approach is valid for both non-relativistic and relativistic approximations. For the sake of generality, we will imply the latter one, considering that all the expressions also apply for non-relativistic or scalar-relativistic Hamiltonians.

The paper is organized as follows. Section~\ref{sec:theory} is focused on the general description of the new method of RDM construction with a further emphasis on the $0h1p$, $0h2p$, and $0h3p$ Fock space sectors. Then, the computational cost of the new method is analyzed and compared with that of solving the FS CC amplitude equations. Section~\ref{sec:pilot} describes a pilot application. The series of contracted ANO-type basis sets for thorium designed specifically for fully relativistic calculations were built by averaging the set of FS CC pure-state density matrices, and the accuracy of such a contraction technique is analyzed. The conclusion (Section~\ref{sec:conclusions}) outlines the developments and future prospects.

\section{Theory}\label{sec:theory}

\subsection{Fock-space coupled cluster method}

The Fock-space formulation of the multireference coupled cluster theory assumes that an exact electronic wavefunction $\ket{\psi_n}$ can be restored from its projection onto the model space $\ket{\tilde{\psi}_n}$ using the exponentially parameterized wave operator $\Omega$ normal-ordered with respect to the common Fermi vacuum~\cite{Lindgren:87,Kaldor:91,Visscher:01,Musial:04,Eliav:Review:22}:
\begin{equation}
\ket{\psi_n} = \Omega \ket{\tilde{\psi}_n}
\quad\quad
\Omega = \{ e^T \},
\label{eq:fscc-ansatz}
\end{equation}
where $T$ stands for the cluster operator and the model vector $\ket{\tilde{\psi}_n}$ is expressed as a linear combination of model-space Slater determinants:
\begin{equation}
\ket{\tilde{\psi}_n} = \sum_\mu c_{n\mu} \ket{\Phi_\mu}
\label{eq:model-vector}
\end{equation}
The wave operator satisfies the intermediate normalization condition (except for the cases when both active holes and particles are involved), while the model vectors are biorthonormalized, i.~e. 
\begin{equation}
\braket{ \tilde{\psi}_n^{\perp\perp} | \tilde{\psi}_m } = \delta_{nm},
\end{equation}
where $\bra{\tilde{\psi}_n^{\perp\perp}}$ stands for a left model vector. Both left and right model vectors are obtained by diagonalization of the effective Hamiltonian $\tilde{H} = (H\Omega)_{cl}$, where the index~\textit{cl} denotes the part of an operator which is \textit{closed} with respect to the model space. The cluster operator $T$ includes contributions with different numbers of annihilation operators acting on active-space spinors. To avoid the problem of redundancy of cluster amplitudes, the amplitude equations are solved successively for the series of model spaces belonging to Hilbert spaces with different numbers of electrons; these spaces are called sectors of the Fock space. The $(n_hh\ n_pp)$ sector comprises model determinants with $n_h$ holes and $n_p$ particles with respect to the chosen closed-shell determinant considered as the Fermi vacuum state. The cluster operator needed to obtain electronic states in the target $(N_hh\ N_pp)$ sector includes contributions from lower Fock space sectors with $n_h \le N_h$ holes and $n_p \le N_p$ particles:
\begin{equation}
T = \sum_{n_h=0}^{N_h} \sum_{n_p=0}^{N_p} T^{(n_hh\ n_pp)}.
\label{eq:cluster-operator}
\end{equation}
Cluster operator $T^{(n_hh\ n_pp)}$ in a given sector consists of operators with different excitation ranks. Within the CCSD approximation considered in the present article, only single and double excitation operators are accounted for:
\begin{equation}
T^{(n_hh\ n_pp)} = T^{(n_hh\ n_pp)}_1 + T^{(n_hh\ n_pp)}_2.
\label{eq:def-ccsd}
\end{equation}
While the CCSD approximation is widely used in practical applications due to its acceptable formal computational cost ($O(M^6)$, where $M$ stands for the number of spinors), for some applications demanding extremely high accuracy the operator (\ref{eq:def-ccsd}) has to be extended with triple excitations (the CCSDT model)~\cite{Musial:04,Oleynichenko:CCSDT:20}.

\subsection{Two general expressions for density matrices}

One-particle \emph{pure-state} reduced density matrix element $\gamma_{qp}^n$ for an electronic state $\ket{\psi_n}$ is defined as follows:
\begin{equation}
\gamma_{qp}^n = \frac{1}{\mathcal{N}_n^2}{\braket{\psi_n|a_p^\dagger a_q|\psi_n}},
\label{eq:def-1-state-dm}
\end{equation}
and a similar formula defines a one-particle \emph{transition} reduced density matrix for the pair of states $\ket{\psi_n}$ and $\ket{\psi_m}$:
\begin{equation}
\gamma_{qp}^{nm} = \frac{1}{\mathcal{N}_n \mathcal{N}_m}\braket{\psi_n|a_p^\dagger a_q|\psi_m},
\label{eq:def-1-tran-dm}
\end{equation}
where the norms $\mathcal{N}_n$, $\mathcal{N}_m$ are given by
\begin{equation}
\mathcal{N}_n = \braket{\psi_n|\psi_n}^{1/2} = \braket{\tilde{\psi}_n|\Omega^\dagger \Omega|\tilde{\psi}_n}^{1/2}.
\label{eq:def:norm-factor}
\end{equation}
The $p$,~$q$,~... indices are used throughout the paper to enumerate spinors regardless of their occupation number in the vacuum determinant, while the $i$,~$j$,~... and $a$,~$b$,~... indices are reserved for hole and particle spinors, respectively. The equations for $\gamma_{qp}^n$ and $\gamma_{qp}^{nm}$ are essentially identical, so we will discuss the latter one and highlight the differences between the pure-state and the transition RDMs if needed.
In the further discussion, we will also use the normal-ordered RDM counterparts, $\left(\gamma^{nm}_N\right)_{qp}$. It is worth mentioning that the \emph{exact} pure-state 1-RDM $\gamma^n$ must be Hermitian, while the transition 1-RDM $\gamma^{nm}$ must be Hermitian conjugate of $\gamma^{mn}$.

The present derivation is based on the formalism of effective operators~\cite{Hurtubise:93} widely used for evaluating expectation and transition values of one-electron properties. There are several ways to define an effective analogue $\tilde{O}$ of a quantum-mechanical operator $O$. The definition guaranteeing the hermiticity of effective analogs of Hermitian operators is given by the relation
\begin{equation}
\braket{\psi_n|O|\psi_m} = \frac{\braket{\tilde{\psi}_n|\tilde{O}|\tilde{\psi}_m}}{\mathcal{N}_n \mathcal{N}_m},
\quad\quad
\tilde{O}=\Omega^\dagger O \Omega.
\label{eq:def:eff-op-herm}
\end{equation}
The other (``non-Hermitian'') definition is more suitable for the theories based on the Bloch formalism and intermediate normalization:
\begin{equation}
\braket{\psi_n|O|\psi_m} =
\frac{\mathcal{N}_n}{\mathcal{N}_m}\braket{\tilde{\psi}_n^{\perp\perp}|\tilde{O}|\tilde{\psi}_m},
\quad
\tilde{O}=\tilde{\Omega} O \Omega,
\label{eq:def:eff-op-non-herm}
\end{equation}
where the mapping $\tilde{\Omega}$ is inverse to the wave operator and is defined through the relation $\bra{\psi_n} =\bra{\tilde{\psi}_n^{\perp\perp}} \tilde{\Omega}$. Assuming the intermediate normalization, one can proceed to the following relation:
\begin{equation}
\braket{\psi_n|O|\psi_m} =
\frac{\mathcal{N}_n}{\mathcal{N}_m}
\braket{\tilde{\psi}_n^{\perp\perp}| (\Omega^\dagger\Omega)^{-1} \Omega^\dagger O \Omega|\tilde{\psi}_m},
\label{eq:def:eff-op-non-herm-2}
\end{equation}
where the inverse of the metric operator $(\Omega^\dagger\Omega)^{-1}$ is restricted to the model space. Normalization factors are naturally canceled for the special case of expectation values ($n = m$). Substituting the operator $\{ a_p^\dagger a_q \}$ into the relations~(\ref{eq:def:eff-op-non-herm}) and~(\ref{eq:def:eff-op-non-herm-2}), one arrives at two working expressions for one-particle reduced density matrices:
\begin{equation}
(\gamma_N)_{pq}^{nm} =
\frac{1}{\mathcal{N}_n \mathcal{N}_m}
\braket{\tilde{\psi}_n|\Omega^\dagger \{ a_p^\dagger a_q \} \Omega|\tilde{\psi}_m},
\label{eq:def:dm-herm}
\end{equation}
\begin{equation}
(\gamma_N)_{pq}^{nm} =
\frac{\mathcal{N}_n}{\mathcal{N}_m}
\braket{\tilde{\psi}_n^{\perp\perp}| (\Omega^\dagger\Omega)^{-1} \Omega^\dagger \{ a_p^\dagger a_q \}  \Omega|\tilde{\psi}_m}.
\label{eq:def:dm-non-herm}
\end{equation}
The substitution of the FS CC Ansatz~(\ref{eq:fscc-ansatz}) into~(\ref{eq:def:dm-herm}) and~(\ref{eq:def:dm-non-herm}) leads to the infinite series including all possible contractions between cluster operators of type $( T^\dagger )^x T^y$. Such series arise both in numerators and denominators. In the special case of the ground state density matrix and the single-reference CC theory the diagrams from the squared norm exactly cancel disconnected terms in the numerator of Eqs.~(\ref{eq:def:eff-op-herm}) and~(\ref{eq:def:dm-herm}), resulting in size-consistent and size-extensive expectation values and density matrices~\cite{Noga:88}; a similar proof is not known for the multireference case. In contrast, the relation~(\ref{eq:def:eff-op-non-herm-2}) can be used to obtain the fully connected expression at least in the second order in $T$~\cite{Zaitsevskii:ThO:23,Oleynichenko:Optics:23} (with some reservations, see below in Section~\ref{sec:conn}). Thus, it seems more beneficial to derive a scheme of constructing RDMs based on the ``non-Hermitian'' expression~(\ref{eq:def:dm-non-herm}).

Approximate pure-state RDMs obtained using Eqs.~(\ref{eq:def:dm-herm}) and~(\ref{eq:def:dm-non-herm}) are Hermitian and non-Hermitian, respectively. The non-hermiticity of pure-state RDMs is typical for CC theories, except for those employing isometric normalization. The diagonalization of such an RDM yields non-coinciding sets of right and left eigenvectors, which can be regarded as natural spinors (and both sets are equally needed to represent a one-particle RDM). This issue seems harmless for the visualization or construction of contracted basis sets. At the same time, the left and right vectors corresponding to the same eigenvalue (the natural occupation number) are ordinarily close to each other. To eliminate this ambiguity, the hermiticity of an RDM can be restored. The situation is similar for transition density matrices. Although transition RDMs are always non-Hermitian by construction, the Hermiticity of the corresponding transition property matrix elements $O_{if} = O_{fi}^*$ is desirable. The second definition of an RDM~(\ref{eq:def:dm-non-herm}) does not fulfill this requirement since it yields a pair of RDMs ($\gamma^{nm},\;\gamma^{mn}$),  violating (normally slightly) the condition $\gamma^{nm}=(\gamma^{mn})^\dag$.
This issue is also well-known in EOM-CC, where transition properties are typically evaluated as a geometric mean between two off-diagonal matrix elements ($\sqrt{O_{if}O_{fi}^*}$)~\cite{Ivanov:19} (the same trick was proposed for FS CC~\cite{Bhattacharya:13,Bhattacharya:14}).

The relations similar to~(\ref{eq:def:dm-herm}) and~(\ref{eq:def:dm-non-herm}) can be obtained for two-particle (transition) reduced density matrices,
\begin{equation}
\gamma^{nm}_{rspq} =
\frac{1}{\mathcal{N}_n \mathcal{N}_m}
\braket{\psi_n| a_p^\dagger a_q^\dagger a_s a_r |\psi_m}.
\label{eq:def-2-rdm}
\end{equation}
Approximate relations for two-particle density matrices can be useful for fast calculations of expectation and transition values of two-body operators. The notable examples include operators characterizing non-trivial interelectronic interactions like the $\mathcal{P}$-odd weak $e-e$ interaction~\cite{Chubukov:17} and hypothetical $\mathcal{T,P}$-odd $e-e$ interaction mediated by the exchange of axion-like particles~\cite{Maison:YbOH:21}. All the conclusions drawn below for the case of one-particle density matrices are also valid for two-particle ones, with only the sets of Brandow diagrams being different.

\subsection{On the connectivity of reduced density matrices}\label{sec:conn}

Now, we derive the working expressions for one-particle reduced density matrices based on Eq.~(\ref{eq:def:dm-non-herm}). In principle, one can insert projection onto the model space $P=\sum_{k} \ket{\tilde{\psi}_k}\bra{\tilde{\psi}^{\perp\perp}_k}$ after the inverse metric operator in~(\ref{eq:def:dm-non-herm}) and perform the inversion of the matrix $\braket{\tilde{\psi}^{\perp\perp}_n | \Omega^\dagger\Omega | \tilde{\psi}_k}$ explicitly~\cite{Zaitsevskii:ThO:23}. Still, the resulting expression contains disconnected diagrams even in the second order in $T$, and how to get rid of them is unclear. An alternative approach represents the inverse metric matrix as a Taylor series. Assuming the exponential Ansatz~(\ref{eq:fscc-ansatz}) for the wave operator, one arrives at
\begin{equation}
(\Omega^\dagger\Omega)_{cl}^{-1} = (1 + (T^\dagger T)_{cl} + ...)^{-1} = 1 - (T^\dagger T)_{cl} + ...
\label{eq:metric-inverse-taylor}
\end{equation}
This expression is valid only for the hole-only ($N_h h\ 0p$) and particle-only ($0 h\ N_p p$) Fock space sectors. In the $1h1p$ sector, the linear closed terms $T_{cl}$ and $T_{cl}^{\dagger}$ arise in~(\ref{eq:metric-inverse-taylor}) due to the lack of intermediate normalization in this sector and the connectivity of the final expression is ruined. The restoration of the intermediate normalization in the $1h1p$ sector is possible~\cite{Zaitsevskii:20}, but this is beyond the scope of this work.

Substitution of the series~(\ref{eq:metric-inverse-taylor}) into the definition~(\ref{eq:def:dm-non-herm}) and retaining only the terms linear and quadratic in $T$ gives an approximate working expression for the one-particle reduced (transition) density matrix:
\begin{align}
& \left(\gamma^{nm}_N\right)_{qp} = 
\frac{\mathcal{N}_n}{\mathcal{N}_m}
\left(\tilde{\gamma}^{nm}_N\right)_{qp} + \nonumber \\
& + \frac{\mathcal{N}_n}{\mathcal{N}_m} \braket{
\tilde{\psi}_n^{\perp\perp}
| T^\dagger \{ a_p^\dagger a_q \} +  \{ a_p^\dagger a_q \} T |
\tilde{\psi}_m
}_{\rm conn} \nonumber \\
& + \frac{\mathcal{N}_n}{\mathcal{N}_m}\bra{\tilde{\psi}_n^{\perp\perp}}
\frac{\{(T^\dagger)^2\}}{2} \{ a_p^\dagger a_q \} +
T^\dagger \{ a_p^\dagger a_q \} T + 
\{ a_p^\dagger a_q \} \frac{\{T^2\}}{2} \nonumber \\
& - (T^\dagger T)_{cl} \{ a_p^\dagger a_q \} \ket{\tilde{\psi}_m}_{\rm conn} ,
\label{eq:1rdm-conn-expr}
\end{align}
where ``conn'' means that only connected Brandow diagrams survive and $\left(\tilde{\gamma}^{nm}_N\right)_{qp}$ stands for the one-particle reduced density matrix confined to the active space (the active-space 1-RDM):
\begin{equation}
\left(\tilde{\gamma}^{nm}_N\right)_{qp} = 
\braket{\tilde{\psi}_n^{\perp\perp}|
\{ a_p^\dagger a_q \}
|\tilde{\psi}_m}.
\label{eq:def-act-1rdm}
\end{equation}
Eq.~(\ref{eq:1rdm-conn-expr}) corresponds to the second-order approximation for the effective one-electron property operator introduced in~\cite{Zaitsevskii:ThO:23,Oleynichenko:Optics:23}. The 1-RDM~(\ref{eq:1rdm-conn-expr}) contracted with a one-electron operator matrix must give exactly the same expectation or transition value of an operator as that obtained within the effective operator formalism. The pilot benchmarks of the latter focused on transition dipole moments show that it possesses an accuracy of a few percent concerning experimental data~\cite{Oleynichenko:Optics:23}; one can expect the same accuracy for density matrices calculated using the expression~(\ref{eq:1rdm-conn-expr}). In principle, one can try to control the accuracy by comparing the results obtained with the second-order expression and its first-order (linear) counterpart; their difference can indicate whether higher-order terms discarded in the present work could be essential.

The first line of Eq.~(\ref{eq:1rdm-conn-expr}) contains active-space 1-RDM. This term can be (very roughly) regarded as responsible for accounting for static correlations (note that the active-space 1-RDM also implicitly accounts for a part of the outer space contributions through the model vector rotations, making such a picture quite approximate). The second and the third lines contain the terms linear and quadratic in cluster amplitudes, respectively. These terms can be interpreted as accounting for dynamic electron correlations by ``dressing'' the active-space 1-RDM. The presence of the renormalization term in the fourth line of~(\ref{eq:1rdm-conn-expr}) is a profound consequence of the intermediate normalization adopted in the FS CC theory. It is worth noting that the renormalization contribution to the 1-RDM (albeit of a slightly different type) arises in the exact expression involving amplitudes of the $\Lambda$-operator~\cite{Szalay:95} (at the lowest order of the many-body perturbation theory $\Lambda \approx T^\dagger$).

It can be shown~\cite{Zaitsevskii:ThO:23} that for the particular case of pure-state density matrices ($n = m$ and the ratio $\mathcal{N}_n/\mathcal{N}_m = 1$) the expression~(\ref{eq:1rdm-conn-expr}) contains only connected diagrams, thus the proper size-extensivity and size-consistency of density matrices and matrix elements of operators is ensured. The situation is not so evident for transition RDMs and properties. In the second order in $T$, normalization factors are represented by connected diagrams, but the whole expression~(\ref{eq:1rdm-conn-expr}) is not guaranteed to be connected if $n \ne m$.
%
%
%
The $\mathcal{N}_n/\mathcal{N}_m$ factor potentially spoiling the connectivity disappears if one considers not the transition density matrix itself but the product of transition matrix elements $|O_{nm}|^2 = O_{nm}O_{mn}^*$. Such a situation is typical; for example, squared absolute values of transition matrix elements drive radiative properties.

There is also the other possibility to get rid of the $\mathcal{N}_n/\mathcal{N}_m$ ratio and proceed to the size-consistent quantities. One of the most important applications of transition RDMs is the analysis of electronic excitation processes in terms of natural transition spinors, NTSs (or orbitals in the non-relativistic picture). Interestingly, one can achieve the exact connectivity of these entities. For this, it is enough to note that NTSs can be obtained through the relations similar to that proposed in~\cite{Martin:NTO:03}:
\begin{equation}
 (\gamma^{nm} \gamma^{mn}) \bm{U} = \bm{\lambda} \bm{U}
\quad\quad
 (\gamma^{mn} \gamma^{nm} ) \bm{V} = \bm{\lambda} \bm{V},
\label{eq:def-nts}
\end{equation}
where $\bm{U}$ and $\bm{V}$ stand for the matrices transforming molecular spinors to initial and final NTSs, respectively, and the $\bm{\lambda}$ is a diagonal matrix containing weights of the corresponding pairs of NTSs. Unlike the commonly used formula based on the singular value decomposition of $\gamma^{nm}$ (see, e.~g., \cite{Krylov:NTO:20}), the relations~(\ref{eq:def-nts}) imply the desired mutual cancellation of the wavefunction norms. 

\subsection{Diagrammatic formulation \\ and formal computational scaling}

\begin{figure}
\includegraphics[width=0.9\columnwidth]{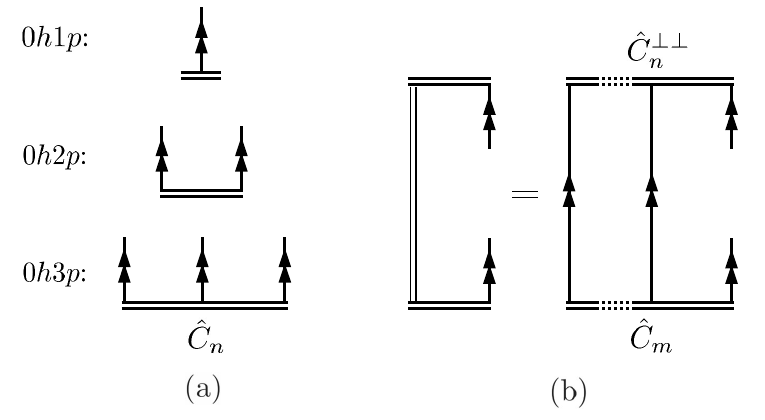}
\caption{Diagrammatic representation of (a) a model vector in the $0h1p$, $0h2p$ and $0h3p$ Fock space sectors; (b) an active-space one-particle reduced density matrix $\tilde{\gamma}_{qp}^{nm}$. Double arrows denote lines corresponding to active spinors. Vertical double line means that some pairs of indices contract model vectors; such a notation allows one to use the same diagram for the active-space 1-RDM in different sectors. Two-particle and three-particle active-space RDMs are defined similarly. The notation is adopted from~\cite{Hanauer:PhD:13}.}
\label{fig:act-dm}
\end{figure}

The working expression~(\ref{eq:1rdm-conn-expr}) explicitly includes coefficients of model vectors. To derive all expressions in a fully diagrammatic manner, it is convenient to introduce an additional second-quantized representation of model vectors $\ket{\tilde{\psi}_n} = \sum_\mu c_{n\mu} \ket{\Phi_\mu}$. For the $N_p$-particle Fock space sector it is written as~\cite{Hanauer:11,Hanauer:PhD:13}
\begin{equation}
\hat{C}_n = \frac{1}{N_p!} \sum_{\mu = u_1, ..., u_{N_p}} 
a^\dagger_{u_1} ... a^\dagger_{u_{N_p}} c_{n\mu} \ket{\Phi_0},
\end{equation}
where $\ket{\Phi_0}$ stands for the Fermi vacuum determinant. Brandow diagrams representing this operator are shown in Fig.~\ref{fig:act-dm}a. Now the arbitrary-rank RDMs confined to the active space ($\tilde{\gamma}^{nm}$) needed to build the ordinary 1-RDMs $\gamma^{nm}$ can also be conveniently represented (Fig.~\ref{fig:act-dm}b). In the present work, we consider the particle-only sectors $0h1p$, $0h2p$, and $0h3p$ as they seem to us the most important from the point of view of practical applications; within the CCSD approximation, the use of higher sectors is impractical due to the abrupt decrease of accuracy caused by the lack of additional cluster operators in these sectors~\cite{Hughes:92}. For the holes-only Fock space sectors, all diagrams are actually the same, but active particle lines must be replaced with active hole lines.

\begin{figure}
\includegraphics[width=0.9\columnwidth]{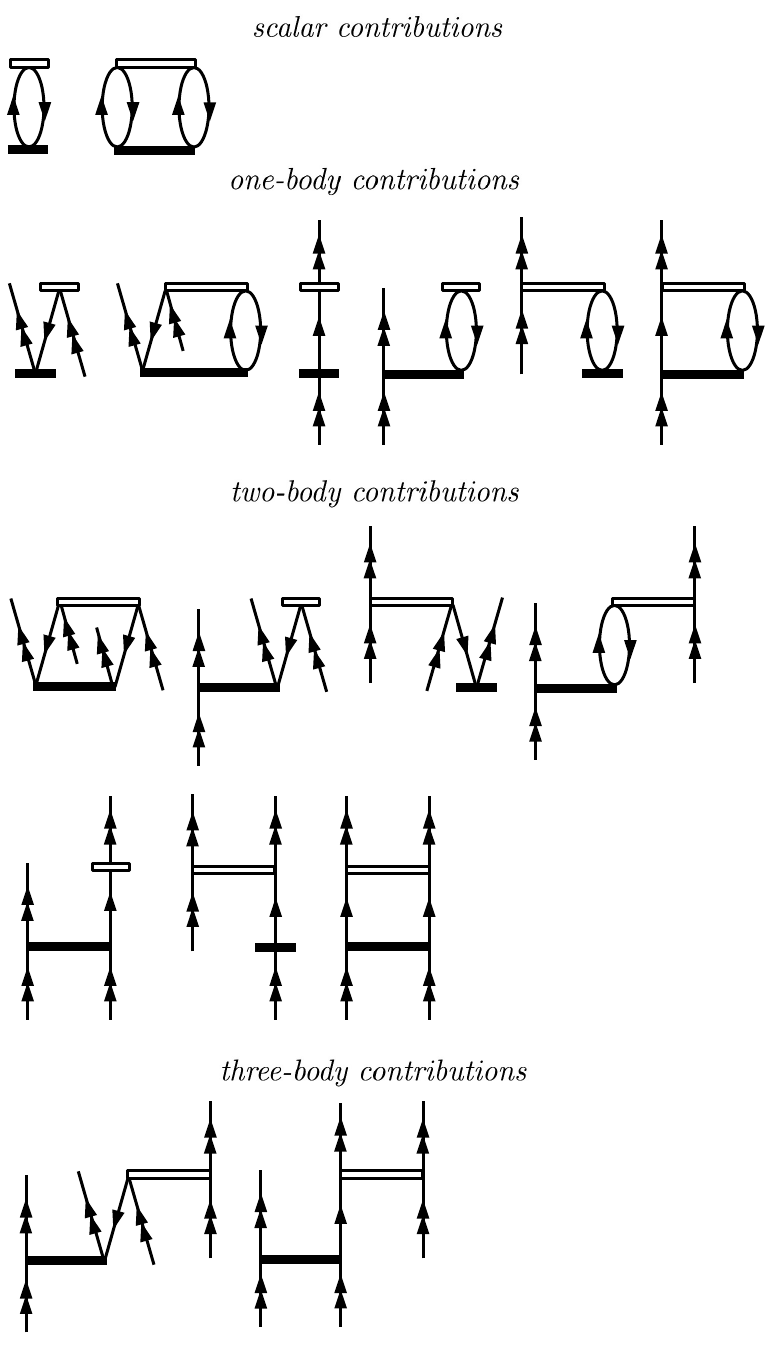}
\caption{Brandow diagrams representing contributions to the metric operator $\Omega^\dagger \approx 1 + T^\dagger T$ in the CCSD approximation and Fock space sectors up to $0h3p$. White and black vertices represent the $T^\dagger$ and $T$ operators, respectively.
}
\label{fig:metric}
\end{figure}

Normalization factors can be evaluated without resorting to active-space reduced density matrices. In the second order in $T$:
\begin{equation}
\mathcal{N}_n^2 = \braket{\tilde{\psi}_n|\Omega^\dagger\Omega|\tilde{\psi}_n} \approx 1 + \braket{\tilde{\psi}_n|(T^\dagger T)_{cl}|\tilde{\psi}_n}.
\label{eq:norm-factor-2nd}
\end{equation}
The term $(T^\dagger T)_{cl}$ can be represented as a sum of the scalar (zero-body), one-body, two-body, etc. contributions (see Fig.~\ref{fig:metric}). The Fock space sector with $N_p$ particles includes up to $N_p$-body terms. In practice, it is convenient to construct Brandow diagrams representing these terms and then use Slater rules to evaluate overlap integrals~(\ref{eq:norm-factor-2nd}).

\begin{figure}
\includegraphics[width=0.9\columnwidth]{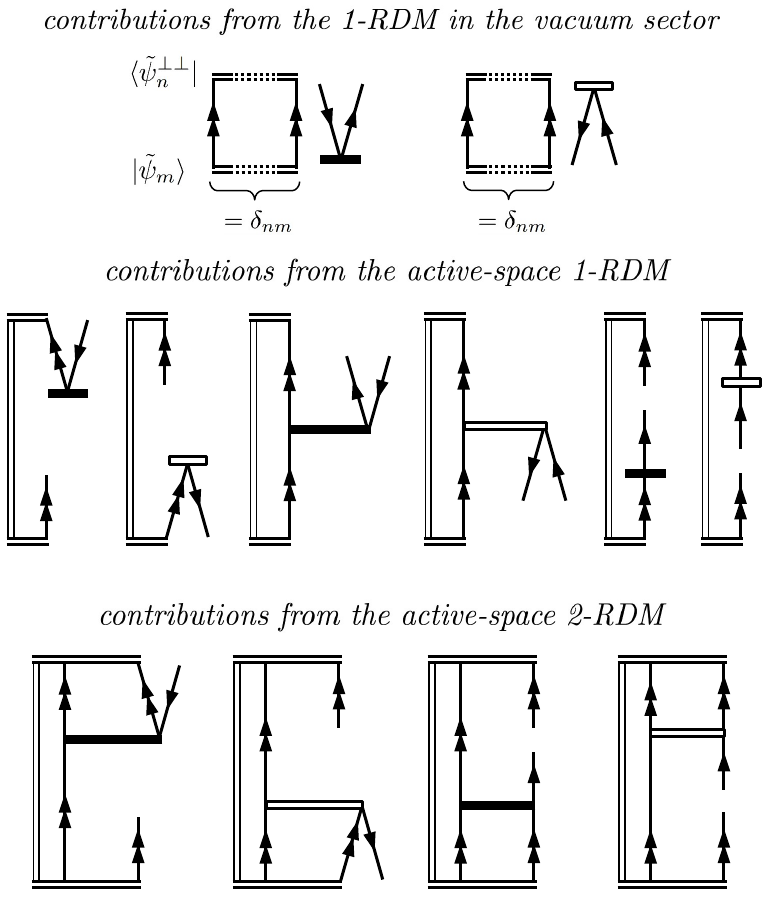}
\caption{Brandow diagrams representing contributions to the one-particle density matrix linear in $T$.}
\label{fig:linear}
\end{figure}

The evaluation of the remaining terms in Eq.~(\ref{eq:1rdm-conn-expr}) is less straightforward and is most conveniently performed using active-space RDMs. The diagrams corresponding to linear terms of this expression are shown in Fig.~\ref{fig:linear}. The vertices depicting the $\{a_p^\dagger a_q\}$ operator are not shown to keep the figure clear, but they are implied and must be taken into account to obtain proper phase factors (see, for example,~\cite{ShavittBartlett:09} for the discussion on analogous diagrams arising in the $0h0p$ sector). There are several important points to be underlined:

(a) the expression for pure-state density matrices contains contributions from the vacuum sector, disappearing for transition density matrices since model vectors are biorthogonal to each other. In principle, one can construct this ``core'' contribution to RDM up to the infinite order in $T$ by simply using the $\Lambda$-equation approach in the $0h0p$ sector while keeping approximate contributions from non-trivial sectors;

(b) the set of diagrams seems to be ``symmetric'' with respect to the $T^\dagger$ and $T$ operators, and the resulting pure-state RDMs at first sight appear Hermitian. This is not so since left and right model vectors are not complex conjugate to each other, and thus the active-space RDMs, as well as their dressed counterparts, are inherently non-Hermitian;

(c) the diagrams list is the same for all Fock space sectors, but for the given $N_p$-particle sector, all active-space RDMs of rank $N_p+1$ or higher are equal to zero, and the corresponding diagrams disappear. For example, in the $0h1p$ sector diagrams, the active-space 2-RDM is zero, and there is no need to account for diagrams containing this vertex (see Fig.~\ref{fig:linear});

(d) in the particular case of the $0h3p$ sector, linear terms are absent within the CCSD approximation and appear only if the triple excitation operator $T_3^{0h3p}$ is explicitly included in the model. Moreover, for the CCSD case, density matrices in the $0h3p$ sector include only the cluster amplitudes from the lower sectors. One can expect an abrupt decrease in the accuracy of calculated RDMs when passing from $0h2p$ to $0h3p$ (in a complete analogy with excitation energies~\cite{Eliav:Review:15}).

The number of possible contractions needed to construct a density matrix using the expression~(\ref{eq:1rdm-conn-expr}) proliferates with the overall power of cluster operators. The second-order approximation has dozens of diagrams, and one can expect hundreds of terms in cubic and quartic approximations. Such a situation clearly appeals to the automatic code generation techniques to obtain and study working expressions for orders higher than the second one.

\begin{table}
\centering
\caption{Formal computational scaling (an asymptotic estimate of the number of floating-point operations) of the second-order method of 1-RDM construction compared to the scaling of one iteration of solving the FS CCSD amplitude equations and construction of the most costly RDM confined to the active space (for the CCSD approximation). $A$, $O$, and $V$ stand for the number of active, occupied, and virtual spinors, respectively.}
\label{tab:cost}
\renewcommand{\arraystretch}{1.5}
\begin{tabular*}{\columnwidth}{l@{\extracolsep{\fill}}cccc}
\hline
\hline
&  $0h0p$   & $0h1p$      & $0h2p$      & $0h3p$   \\
\hline
amplitude equations &  $O^2V^4$ & $A^1O^1V^4$ & $A^2V^4$ & $A^6V^1$\fn1 \\
active-space RDMs &   $-$        & $A^2$       & $A^4$       & $A^6$    \\
one-particle RDM &  $O^2V^3$ & $A^1O^1V^3$ & $A^3O^1V^2$ & $A^6V^1$ \\
\hline
\hline
\end{tabular*}
\footnotetext[1]{For the non-iterative construction of an effective Hamiltonian. Within the FS CCSD method, one cannot define cluster amplitudes specific for the $0h3p$ sector, and thus amplitude equations are absent~\cite{Hughes:92,Meissner:0h3p:20}.}
\end{table}

Table~\ref{tab:cost} summarizes the formal computational scaling of the one-particle density matrix construction compared to solving amplitude equations and constructing active-space RDMs. It can be seen that for the low sectors ($0h0p$ -- $0h2p$), the number of floating-point operations is always lower than the cost of one iteration of the CCSD amplitude equations, provided that the numbers of active ($A$) and virtual $(V)$ spinors are related as $A << V$. In the $0h3p$ sector, the number of operations needed to construct an effective Hamiltonian and a 1-RDM are comparable.

The new method of density matrix construction was implemented as a computer code for the $0h0p$ -- $0h3p$ Fock space sectors and included in the {\sc exp-t} program package~\cite{Oleynichenko:EXPT:20,EXPT:23}.

\bigskip
\bigskip

\section{Pilot application: \\ contracted basis sets for relativistic calculations of excited states}\label{sec:pilot}

\subsection{ANO-type contracted basis sets}

Highly accurate calculations of excited electronic states of systems containing heavy elements typically require the construction of special atomic basis sets explicitly designed for a problem under consideration. 
The most widely used basis sets well-suitable for coupled cluster calculations like Dyall's basis sets~\cite{Dyall:06,Dyall:12,Dyall:24}, ANO-RCC~\cite{Almlof:91,Widmark:90,Roos:05} and (aug-)cc-pVXZ-DK3/X2C~\cite{Peterson:15,Feng:17} sets perform great for modeling of heavy-element atoms and compounds in their ground states, but sometimes can possess unsatisfactory accuracy in excited-state calculations~\cite{AthanasakisRaFPinning:2023}, especially in cases where the number of $d$- or $f$-electrons changes upon excitation or chemical processes (such a situation is typical for lanthanide and actinide compounds). The practical experience~\cite{AthanasakisRaFPinning:2023} shows that the common deficiency of the mentioned sets is the lack of functions with high angular momenta, first of all, $h$- and $i$-functions, which are of crucial importance for even a qualitatively correct description of electronic states involving open $d$- and $f$-shells. In the particular case of atoms and highly-symmetric small molecules, one can augment a basis set with primitive Gaussians of these types and optimize their exponential parameters. This solution has become too computationally demanding already for small molecules containing three or more atoms, and it is prohibitive for medium-size systems like complex compounds or cluster models used to describe impurity centers and/or local processes in solids~\cite{Oleynichenko:YPO4:24}.

In the present work, we construct ANO-type contracted basis sets for relativistic calculations of multiple electronic states and spectra of heavy-element compounds using appropriate averaging of approximate FS RCC 1-RDMs. The procedure generally follows the idea of Widmark, Roos, and co-workers~\cite{Widmark:90,Roos:05}, but treats scalar-relativistic and spin-dependent relativistic effects on an equal footing, employs a higher-level method to solve correlation problem and is designed for a well-balanced treatment of the ground and numerous excited states.

In our previous works~\cite{Skripnikov:15a,Skripnikov:15c,Skripnikov:ThO:16} the ANO-like compact basis set generation procedure, as adapted in Ref.~\cite{Skripnikov:E120:13}, employed averaging of 1-RDMs obtained from scalar-relativistic single-reference CC calculations of atoms and molecules in electronic states relevant to the problem under consideration. We employ the atomic orbital (AO) basis set representation of a density matrix to obtain contraction coefficients of a natural basis set. In the scalar-relativistic case, density matrix elements between molecular (atomic) spin-orbitals with different spin projections equal zero. Thus, on an AO basis, the full one-particle density matrix has only two blocks corresponding to the ``diagonal'' spin-up or spin-down cases, and one averages over these two blocks. Additionally, one averages over all projections of AO basis functions. The resulting set of contracted basis functions is sorted in descending order by their natural populations. Within this scheme, contracted basis functions necessary to describe changes in atomic functions due to the spin-orbit interaction, such as functions representing the difference of radial parts of the $np_{3/2}$ and $np_{1/2}$, $nd_{5/2}$ and $nd_{3/2}$, etc. spinors, were obtained in the relativistic atomic Hartree-Fock calculation and then manually added to the basis set. The latter step can potentially lead to linear dependencies. Moreover, the resulting basis set becomes less compact than it could be. The latter circumstance can be critical for coupled cluster calculations accounting for higher excitations (for example, CCSDT).

This approach has been recently extended by including in the averaging procedure not only the density matrices obtained in the scalar-relativistic CC calculations but also the relativistic density matrices calculated at the two-component Hartree-Fock level (i.~e., the density matrix due to particular occupied and virtual Dirac-Hartree-Fock spinors) for all of the relevant states~\cite{AthanasakisRaFPinning:2023}. For this, it was necessary to add contributions to the AO density matrix induced by the terms such as $\gamma_{pp}=1$ in the basis of atomic or molecular occupied or virtual two-component complex spinors $\phi_p = \sum_\mu (C^\alpha_{p,\mu}\chi_{\mu} \alpha + C^\beta_{p,\mu}\chi_{\mu} \beta)$. Here, $\chi_{\mu}$ are atomic orbitals (AOs), $\alpha$, $\beta$ are spin functions and $C^{\alpha/\beta}_{p,\mu}$ are expansion coefficients. Due to the spinor structure, nonzero contributions to the AO density matrix 
of the type $C^\alpha_{p,\mu} C^{\beta*}_{p,\nu}$
(``mixed-spin'' blocks)
can appear (see, for example,~\cite{Wullen:10}). In the procedure used, we averaged over both spin-diagonal (as in the scalar-relativistic case) and mixed-spin blocks. We also averaged over the real and imaginary parts of the density matrix. Such a procedure prevents one from manually adding the functions required to describe spin-orbit effects. If the scalar-relativistic correlated density matrices are averaged with the density matrices, which include pairs of spin-orbit-split functions $\phi_p$, the mentioned linear dependence problem disappears, and a more compact description of the basis set is possible. The basis set for Ra constructed in~\cite{AthanasakisRaFPinning:2023} was compact and accurate enough to describe multiple electronic states of the RaF molecule at the relativistic FS CCSDT level~\cite{AthanasakisRaFPinning:2023}. In practice, one can ensure that the functions required to describe spin-orbit splitting accurately are included in the basis set by controlling the natural occupations at the final ANO selection step. However, in this case the interference between spin-orbit and correlation effects is ignored. This factor seems unimportant for $g$-, $h$- and $i$-functions but can be notable for basis functions with lower angular momenta. Therefore, an alternative approach is to consider only high-angular momentum ANO-type basis functions while using primitive basis sets for low-angular momenta~\cite{Oleynichenko:CCSDT:20,Skripnikov:2020c}.

The next logical step is the averaging over a series of density matrices for different electronic configurations and/or excited states of a heavy ion or an atom obtained within some version of the fully relativistic coupled cluster method. This idea was employed in a recent work~\cite{Oleynichenko:YPO4:24}, where the fully relativistic single-reference CCSD method was employed to generate 1-RDMs. However, the single-reference CC method can treat only electronic states with pronounced single-determinant character, e.~g. high-spin states, resulting in a contracted basis set biased towards such states. Thus, it is reasonable to use a multi-state multireference coupled cluster method like FS CC to generate density matrices for the whole manifold of the states of interest on equal footing regardless of their determinantal structure. Note that such density matrices are not expected to be very accurate. Within such an approach, both correlation and spin-orbit effects are considered on an equal footing. For this purpose, the {\sc natbas} code~\cite{Skripnikov:E120:13,AthanasakisRaFPinning:2023} was extended to work with density matrices generated by the relativistic FS CC method.

\subsection{Basis set for thorium}

We consider constructing the ANO-type basis set for thorium as a pilot application of the finite-order method to calculate approximate RDMs. We have chosen Th due to the importance of its compounds ThO and ThF$^+$ for the searches of the physics beyond the Standard model~\cite{Gresh:ThF:16,Andreev:ThO:18,ArrowsmithKron:24}, increasing interest in Th compounds due to the thorium fuel cycle and possible use in the new generation of frequency standards~\cite{Beeks:21}. Moreover, unlike many other lanthanides and actinides, the electronic states of thorium ions possess only a few unpaired electrons and can then be reached using the relativistic FS CC method.

In the present paper, we focus on the construction of a basis set adapted to calculations with the highly accurate non-local generalized relativistic pseudopotential (GRPP) of Titov, Mosyagin et al.~\cite{Titov:99,Petrov_2004,Mosyagin:16,Oleynichenko:LIBGRPP:23}. The pseudopotentials of this series not only effectively account for the scalar-relativistic effects and spin-orbit interaction but also absorb the bulk of effects arising from the two-electron Breit interaction, the finite size of a nucleus, electron self-energy, and vacuum polarization effects due to the use of the model QED potential~\cite{Shabaev:13} at the GRPP construction stage. Uncertainties arising from the use of the pseudopotential approximation for Th was previously shown to be not exceeding 50~cm$^{-1}$~\cite{Oleynichenko:LIBGRPP:23}; note that the Breit contribution not accounted for in many routine four-component Dirac-Coulomb calculations can be an order of magnitude larger. The pseudopotential employed in the present work excludes 28 core electrons of Th (shells with the principal quantum numbers $n=1-3$).

The set of primitive Gaussians was based on that of the dyall.ae4z basis set~\cite{Dyall:06,Dyall:12}. For the $s$-, $p$-, $d$- and $f$-blocks, the five tightest primitive functions corresponding to the outercore region of an atom were re-optimized in the series of SCF calculations of the Th$^{4+}$ ion to fit the behavior of outercore pseudospinors better. Moreover, the original $i$-functions were replaced with 5 primitive Gaussians with exponential parameters $\zeta_n = 0.6 \times 2^{n-1}$, $n=1-5$. The resulting primitive basis set has the composition $(17s17p15d14f8g7h5i)$. It was found to be saturated in the sense that excitation energies calculated by the FS CCSD method for chosen electronic configurations of Th$^{2+}$ (see below) were found to be stable within 5~cm$^{-1}$ with respect to the further extension of this basis set.

It is worth noting that very accurate predictions of excitation spectra of actinide compounds also require $k$-functions. For example, we found that the corresponding contributions to excitation energies of Th$^{2+}$ can reach ca.~400~cm$^{-1}$. Moreover, due to the very dense spectrum of low-lying states the correct order of the ground ($5f6d\ ^3H^o_4$) and the first excited ($6d^2\ ^3F_2$ at 63~cm$^{-1}$) states cannot be reproduced without the inclusion of two $k$-functions with exponential parameters 1.98 and 3.96 (see also~\cite{Eliav:Th:02}). Due to the current limitations of the {\sc dirac} program package, one cannot use $k$-functions in pseudopotential calculations, and thus, $k$-functions were not included in the final basis set. If needed, corrections for $k$-functions can be obtained from a four-component Dirac-Coulomb(-Gaunt) calculation within the exact two-component molecular mean field (X2C MMF) approximation~\cite{Sikkema:x2cmmf:09}. Moreover, the problems arising from the lack of $k$-functions are typical for atomic calculations of $f$-elements and even for 7$s$-elements~\cite{Skripnikov:2021a}. Still, they are not so keen for molecules~\cite{Skripnikov:2021a} (in turn, pseudopotentials are more widely used in molecular calculations).

The choice of particular pure-state density matrices to be averaged strongly depends on electronic structure peculiarities and a manifold of target states of a particular system. In most thorium compounds, its oxidation state lies between +1 and +2. For example, electronic states of the ThF$^+$ molecular ion lying in the optical range can be roughly regarded as the states of Th$^{2+}$ split by the field of the F$^-$ anion. For this reason, the present version of the ANO-type basis set was constructed for the set of 481 pure-state density matrices obtained by the relativistic FS CCSD method for the electronic states of Th$^{2+}$ lying below 70000~cm$^{-1}$ and corresponding to the electronic configurations $5f6d$, $6d^2$, $5f7s$, $6d7s$, $7s^2$, $5f^2$, $5f7p$, $6d7p$, $7s7p$ (the full list of calculated energy levels compared with experimental data can be found in the Supplementary Materials). The set of valence states of Th and/or configurations can be further tuned for any particular problem. The closed-shell ground state of Th$^{4+}$ was considered as a Fermi vacuum, and the target electronic states were obtained in the $0h2p$ Fock space sector. The active space comprised $7s$, $7p$, $6d$ and $5f$ spinors or Th$^{4+}$. To avoid the intruder-state problem, the intermediate Hamiltonian for the incomplete main model spaces (IH-IMMS) technique~\cite{Zaitsevskii:QED:22} was applied, and the $7p^2$ states potentially dangerous for convergence were considered as belonging to the intermediate model space. All 60 electrons of Th$^{2+}$ not excluded by the pseudopotential were correlated, and the energy cutoff for the one-electron spectrum was set to +200~a.~u. The two-component relativistic Hartree-Fock calculations and the subsequent transformation of molecular integrals were carried out using the {\sc dirac} program package~\cite{DIRAC_code:19,Saue:20} supplied by the {\sc libgrpp} library~\cite{Oleynichenko:LIBGRPP:23} to calculate integrals over the GRPP operator. The {\sc exp-t} program~\cite{Oleynichenko:EXPT:20,EXPT:23} was used to solve the FS CCSD amplitude equations and construct density matrices.

\begin{table}
\centering
\caption{Comparison of basis sets for thorium constructed in the present work. The Th$^{2+}$ excitation energies were calculated with different contracted basis sets and compared to their counterparts obtained for the primitive version of the basis. Maximum absolute errors and mean absolute deviations with respect to the calculation with the primitive basis are given in units of cm$^{-1}$.}
\label{tab:basis}
\renewcommand{\arraystretch}{1.5}
\begin{tabular*}{\columnwidth}{l@{\extracolsep{\fill}}lccc}
\hline
\hline
\multicolumn{2}{c}{contraction scheme} & size\fn1 & max abs & mean abs \\
[-1ex] $spdf$ part & $ghi$ part  &          & error   & deviation \\
\hline
$17s17p15d14f$ & $8g7h5i$ & 455 & $-$ & $-$ \\
[1ex] $17s17p15d14f$ & $6g5h3i$ & 389 & 160  & 48  \\
$9s10p9d8f$ & $6g5h3i$ & 288  & 674  & 63  \\
[1ex] $17s17p15d14f$ & $5g4h2i$ & 356 & 423  & 129 \\
$8s9p8d7f$ & $5g4h2i$   & 239 & 1354 & 144 \\
[1ex] $17s17p15d14f$ & $4g3h1i$ & 323 & 936  & 327 \\
$8s8p7d6f$ & $4g3h1i$    & 191 & 2110 & 449 \\
\hline
\hline
\end{tabular*}
\footnotetext[1]{Total number of spherical Gaussian basis functions.}
\end{table}

\begin{figure*}
\includegraphics[width=0.9\textwidth]{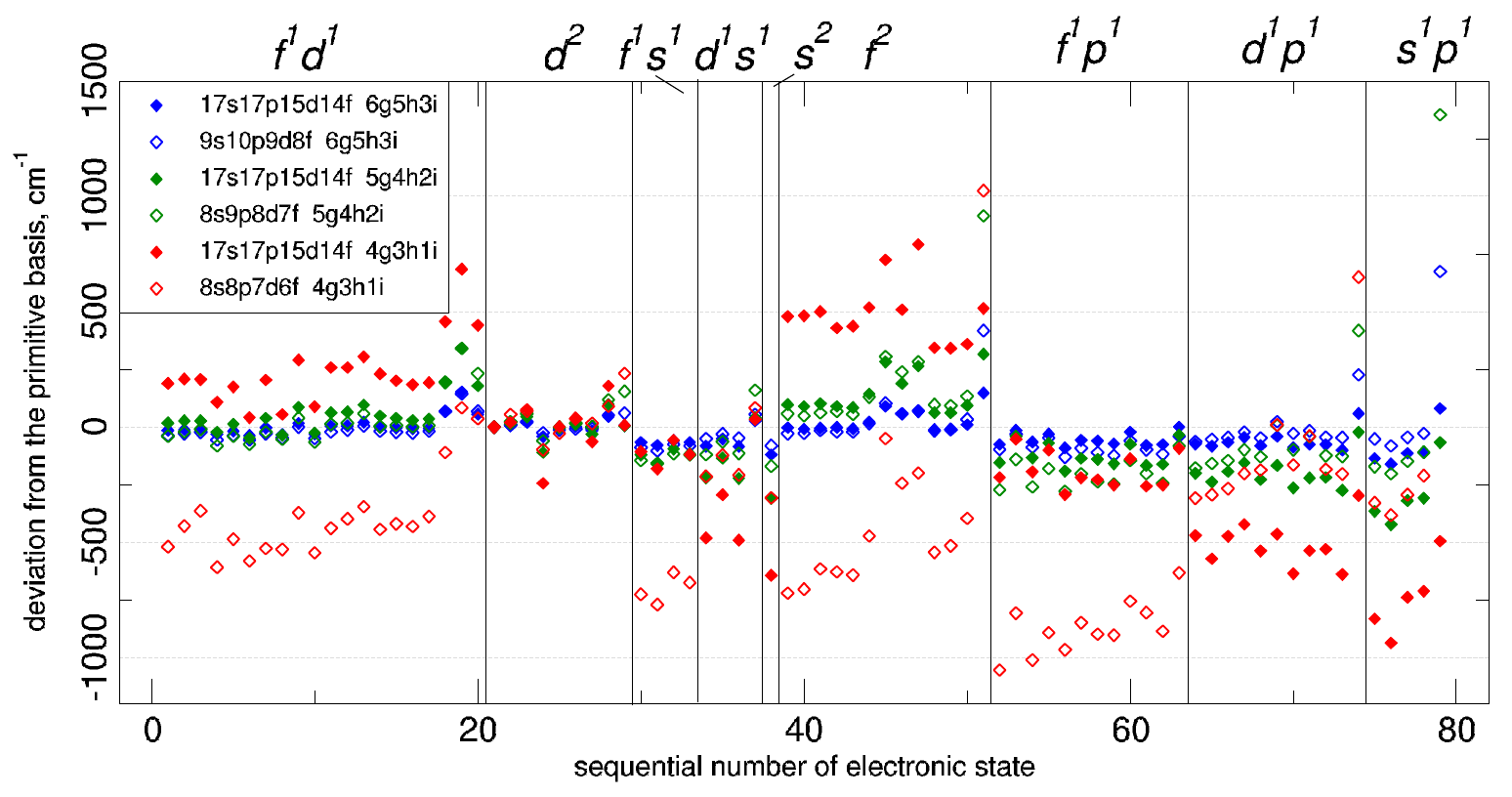}
\caption{Deviations of excitation energies of the Th$^{2+}$ ion calculated using different ANO-type contracted basis sets from those calculated using the primitive set $(17s17p15d14f8g7h5i)$.}
\label{fig:thorium}
\end{figure*}

Two series of contracted Gaussian basis sets were constructed based on ANO occupation numbers. The first one included three basis sets with a common fully uncontracted $spdf$ part but different high angular momentum parts, namely, $[6g5h3i]$, $[5g4h2i]$ and $[4g3h1i]$. The second series was designed to be more compact and thus suitable for computationally demanding molecular calculations. It included three sets with the same $ghi$ parts and contracted functions with lower angular momenta. Compositions of the constructed basis sets and their performance are summarized and compared in Tab.~\ref{tab:basis}; basis sets are also available in the Supplementary Materials.

\subsection{Results and discussion}

Figure~\ref{fig:thorium} shows the deviations of excitation energies of the Th$^{2+}$ ion calculated using different contracted basis sets from those obtained with the primitive basis set. To make trends more explicit, electronic states were rearranged and grouped by their leading configurations (however, the pattern of appearance of these configurations in the experimental spectrum is still approximately preserved). The lowest deviations from the primitive set are observed for the $6d^2$ states, since FS CCSD predicts the $6d^2\ ^3F_2$ state to be the ground one, and all excitation energies are calculated with respect to this level. The following trends can be identified:

(a) deviations are well-balanced for the whole manifold of electronic states considered for the first two schemes of contraction of the high angular momentum part of a basis, $[6g5h3i]$ and $[5g4h2i]$. Deviations increase dramatically when passing to $[4g3h1i]$; for the $5f^2$, $6d7p$, and $7s7p$ states the deviations are approximately three times larger than for the $5f6d$, $6d^2$ and $5f7s$ states. Such a situation can lead to imbalanced and hardly predictable uncertainties in molecular calculations with these basis sets;

(b) contraction of the ``low'' angular momenta ($spdf$) part does not actually decrease accuracy for the first two contracted sets (except for the high-energy singlet state $7s7p\ ^1P^o_1$). If an accurate description of such states is needed, an additional optimized primitive diffuse function can be added. The quality of contraction is significantly lost for the most compact $[8s8p7d6f4g3h1i]$ set. This is because there is a too stringent limit on natural occupation numbers, resulting in an exclusion of functions needed to accurately describe differences in radial parts of spinors induced by spin-orbit splitting. It can be recommended that functions with low angular momenta be excluded one by one, regardless of their natural populations, to avoid such losses in accuracy. Thus, the contraction of the ``low'' angular momenta part is obviously more beneficial than the contraction of the ``polarization + correlation'' part ($ghi$-functions). This also gives an advantage in the final number of basis functions;

(c) the largest deviations correspond to electronic states, which can be classified as spin singlets (for example, three rightmost states in the $5f6d$ manifold are $^1F^o$, $^1H^o$ and $^1P^o$). This is a well-known issue due to strong dynamic correlations in these states, which are poorly described by electronic wavefunctions composed of Slater determinants. The problem is aggravated by the fact that there are few such states in the spectrum, and the averaging procedure is biased towards triplet states. This problem, in principle, can be solved (at least partially) by increasing the weights of such states during the averaging step.

Finally, we note that such a striking loss of accuracy for the contracted sets of type $[4g3h1i]$ can be significantly leveled out in molecular calculations due to basis functions centered on neighboring atoms. Moreover, such compact basis sets still can be very useful to calculate corrections for contributions of triple (and higher) excitations using the very computationally demanding FS CCSDT or CCSDT(Q) methods~\cite{Oleynichenko:CCSDT:20,Zaitsevskii:QED:22,Zaitsevskii:RaF:22,AthanasakisRaFPinning:2023,Skripnikov:E120:13,Skripnikov:15a,Skripnikov:15c,Skripnikov:ThO:16}.

\section{Conclusions}\label{sec:conclusions}

An efficient finite-order method has been proposed and implemented to calculate approximate pure-state and transition density matrices within the Fock-space multireference coupled cluster theory. It follows the same logic as the finite-order method of calculation of expectation values and transition moments of one-electron operators developed recently~\cite{Zaitsevskii:ThO:23,Oleynichenko:Optics:23}. The method is cheap in terms of its computational cost and relatively easy-to-implement. The underlying approach can also be applied to other multistate MR CC formulations.

Further development of the described scheme can include accounting for higher excitations and/or contributions cubic or quartic in $T$ (or at least the most important terms of this type). However, the value of the proposed method lies in its low computational cost, while relatively accurate RDMs are still obtained. Thus possible extensions should keep computational complexity lower than that of solving of CC amplitude equations, otherwise the exact method based on $\Lambda$-equations would be preferable.

Fast and rather accurate calculations of density matrices for multiple electronic states reveal possibilities for a systematic generation of contracted Gaussian basis sets oriented at fully relativistic calculations of excited states and spectra of heavy-element compounds. The series of contracted basis sets for thorium was constructed as a pilot application. The other promising field uses natural spinors with significant occupation numbers obtained by FS CCSD as a compact one-electron basis in FS CCSDT calculations~\cite{Haldar:21,Chamoli:24}; the work is currently in progress.

\section{Acknowledgements}

Electronic structure calculations have been carried out using computing resources of the federal collective usage center Complex for Simulation and Data Processing for Mega-science Facilities at National Research Centre ``Kurchatov Institute'', http://ckp.nrcki.ru/.


The work of A.V.O. and A.Z. at NRC ``Kurchatov Institute'' -- PNPI on the new method to construct density matrices, its first program implementation in the {\sc exp-t} package and the generation of the basis sets for Th was supported by the Russian Science Foundation under grant no.~20-13-00225, \url{https://rscf.ru/project/23-13-45028/}. The work of L.V.S. at NRC ``Kurchatov Institute'' -- PNPI on the development of the interface of the {\sc natbas} code to the {\sc exp-t} package was supported by the Foundation for the Advancement of Theoretical Physics and Mathematics ``BASIS'' Grant according to Projects No. 24-1-1-36-1.

\bibliography{bibfile}

\end{document}


\title{Supplementary Materials for: \\ ``Finite-order method to calculate approximate density matrices\\ in the Fock-space multireference coupled cluster theory''}

\author{Alexander~V.~Oleynichenko}
\email{oleynichenko\_av@pnpi.nrcki.ru}

\affiliation{B. P. Konstantinov Petersburg Nuclear Physics Institute of National Research Center ``Kurchatov Institute'' (NRC ``Kurchatov Institute'' -- PNPI), Gatchina, Leningrad district 188300, Russia}

\affiliation{Moscow Institute of Physics and Technologies (National Research University), Institutskij pereulok 9, Dolgoprudny, Moscow region 141700, Russia}

\author{Andr\'ei~Zaitsevskii}
\email{zaitsevskii\_av@pnpi.nrcki.ru}

\affiliation{B. P. Konstantinov Petersburg Nuclear Physics Institute of National Research Center ``Kurchatov Institute'' (NRC ``Kurchatov Institute'' -- PNPI), Gatchina, Leningrad district 188300, Russia}

\affiliation{Department of Chemistry, M.~V. Lomonosov Moscow State University, 119991 Moscow, Russia}

\author{Leonid~V.~Skripnikov}
\email{skripnikov\_lv@pnpi.nrcki.ru}

\affiliation{B. P. Konstantinov Petersburg Nuclear Physics Institute of National Research Center ``Kurchatov Institute'' (NRC ``Kurchatov Institute'' -- PNPI), Gatchina, Leningrad district 188300, Russia}

\affiliation{Saint Petersburg State University, 7/9 Universitetskaya nab., 199034 St. Petersburg, Russia}

\author{Ephraim~Eliav}
\email{ephraim@tau.ac.il}

\affiliation{School of Chemistry, Tel Aviv University, Tel Aviv 6997801, Israel}

\date{\today}

\maketitle

\onecolumngrid

\begin{table}[H]
\centering
\renewcommand\thetable{S1}
\caption{Energy levels (cm$^{-1}$) of the Th$^{2+}$ cation calculated by the relativistic FS CCSD method using the primitive Gaussian basis set $(17s17p15d14f8g7h5i)$ compared to the experimental data~\cite{Blaise:92}.}
\renewcommand{\arraystretch}{1.2}
\begin{tabular*}{\columnwidth}{l@{\extracolsep{\fill}}lcccc}
\hline
\hline
\multicolumn{3}{c}{Th$^{2+}$ energy levels} & Experiment & FS CCSD & Deviation \\
\cmidrule{1-3}
Configuration & Term & $J$ \\
\hline
$5f6d$ & $^3H^o$ & 4 & 0 & 216 & 216 \\
       &        & 5 & 4490 & 4761 & 271 \\
       &        & 6 & 8437 & 8704 & 267 \\
\\
$6d^2$ & $^3F$  & 2 & 63 & 0 & -63 \\
       &        & 3 & 4056 & 4028 & -28 \\
       &        & 4 & 6538 & 6480 & -58 \\
\\
$5f6d$ & $^3F^o$ & 2 & 511 & 842 & 332 \\
\\
$5f7s$ & $^o$ & 3 & 2527 & 3069 & 542 \\
\\
$5f7s$ & $^3F^o$ & 2 & 3182 & 3733 & 551 \\
       &         & 4 & 6311 & 6688 & 377 \\
\\
$5f6d$ & $^o$ & 4 & 3188 & 3421 & 233 \\
\\
$6d^2$ &      & 2 & 4676 & 4590 & -87 \\
\\
$5f6d$ & $^o$ & 3 & 4827 & 5420 & 594 \\
\\
$5f6d$ & $^3G^o$ & 3 & 5061 & 6390 & 1330 \\
       &         & 4 & 8142 & 9121 & 979 \\
       &         & 5 & 11277 & 12730 & 1453 \\
\\
$6d^2$ & $^3P$   & 0 & 5090 & 5039 & -51 \\
       &         & 1 & 7876 & 7861 & -15 \\
       &         & 2 & 10440 & 10443 & 3 \\
\hline
\hline
\end{tabular*}
\end{table}

\begin{table}[H]
\centering
\renewcommand\thetable{S1 (Continue)}
\caption{Energy levels (cm$^{-1}$) of the Th$^{2+}$ cation calculated by the relativistic FS CCSD method using the primitive Gaussian basis set $(17s17p15d14f8g7h5i)$ compared to the experimental data~\cite{Blaise:92}.}
\renewcommand{\arraystretch}{1.2}
\begin{tabular*}{\columnwidth}{l@{\extracolsep{\fill}}lcccc}
\hline
\hline
\multicolumn{3}{c}{Th$^{2+}$ energy levels} & Experiment & FS CCSD & Deviation \\
\cmidrule{1-3}
Configuration & Term & $J$ \\
\hline
$6d7s$ & $^3D$ & 1 & 5524 & 5422 & -102 \\
       &       & 2 & 7176 & 7095 & -81 \\
       &       & 3 & 9954 & 9851 & -103 \\
\\
$5f6d$ & $^o$ & 2 & 6288 & 6571 & 283 \\
\\
$5f7s$ & $^o$ & 3 & 7501 & 8206 & 705 \\
\\
$5f6d$ & $^3D^o$ & 1 & 7921 & 9164 & 1243 \\
       &         & 2 & 10181 & 11308 & 1128 \\
       &         & 3 & 10741 & 12058 & 1317 \\
\\
$5f6d$ & $^o$ & 4 & 8981 & 9977 & 997 \\
\\
$6d^2$ & $^1G$ & 4 & 10543 & 10432 & -111 \\
\\
$5f6d$ & $^3P^o$ & 1 & 11123 & 12324 & 1201 \\
       &         & 0 & 11233 & 12384 & 1152 \\
       &         & 2 &  13208 & 14158 & 949 \\
\\
$7s^2$ & $^1S$ & 0 & 11961 & 11721 & -240 \\
\\
$5f^2$ & $^3H$ & 4 & 15149 & 16454 & 1306 \\
       &       & 5 & 17887 & 19113 & 1226 \\
       &       & 6 & 20771 & 21913 & 1142 \\
\\
$5f6d$ & $^1F^o$ & 3 & 15453 & 17138 & 1685 \\
\\
$6d7s$ & $^1D$ & 2 & 16038 & 16122 & 84 \\
\\
$5f^2$ & $^3F$ & 2 & 18864 & 20384 & 1520 \\
       &    & 3 & 20840 & 22352 & 1511 \\
       &    & 4 & 21784 & 23364 & 1580 \\
\\
$6d^2$ & & 0 & 18993 & 21281 & 2288 \\
\\
$5f6d$ & $^1H^o$ & 5 & 19010 & 21540 & 2530 \\
\\
$5f6d$ & $^1P^o$ & 1 & 20711 & 22890 & 2179 \\
\\
$5f^2$ & $^1G$ & 4 & 25972 & 27620 & 1648 \\
\\
$5f^2$ & $^1D$ & 2 & 28233 & 30138 & 1906 \\
\\
$5f^2$ & $^1I$ & 6 & 28350 & 29376 & 1026 \\
\\
$5f^2$ & $^3P$ & 0 & 29300 & 31337 & 2038 \\
       &       & 1 & 30403 & 32377 & 1975 \\
       &       & 2 & 32867 & 34750 & 1883 \\
\hline
\hline
\end{tabular*}
\end{table}

\newpage

\begin{table}[H]
\centering
\renewcommand\thetable{S1 (Continue)}
\caption{Energy levels (cm$^{-1}$) of the Th$^{2+}$ cation calculated by the relativistic FS CCSD method using the primitive Gaussian basis set $(17s17p15d14f8g7h5i)$ compared to the experimental data~\cite{Blaise:92}.}
\renewcommand{\arraystretch}{1.2}
\begin{tabular*}{\columnwidth}{l@{\extracolsep{\fill}}lcccc}
\hline
\hline
\multicolumn{3}{c}{Th$^{2+}$ energy levels} & Experiment & FS CCSD & Deviation \\
\cmidrule{1-3}
Configuration & Term & $J$ \\
\hline
$5f7p$ & (5/2,1/2) & 3 & 33562 & 34573 & 1011 \\
       &           & 2 & 34996 & 36374 & 1378 \\
\\
$6d7p$ & (3/2,1/2)$^o$ & 2 & 37280 & 37465 & 185 \\
       &            & 1 & 39281 & 39524 & 243 \\
\\
$5f7p$ & (7/2,1/2) & 3 & 38432 & 39537 & 1105 \\
       &           & 4 & 38581 & 39728 & 1147 \\
\\
$7s7p$ & $^3P^o$ & 0 & 42260 & 41962 & -297 \\
       &         & 1 & 45064 & 45143 & 79 \\
       &         & 2 & 55399 & 55807 & 407 \\
\\
$5f7p$ & (5/2,3/2) & 3 & 42313 & 43513 & 1200 \\
       &           & 4 & 43702 & 45082 & 1381 \\
       &           & 2 & 43759 & 45293 & 1534 \\
       &           & 1 & 44603 & 46189 & 1586 \\
\\
$6d7p$ & (5/2,1/2)$^o$ & 2 & 44088 & 44428 & 340 \\
       &            & 3 & 44465 & 44904 & 439 \\
\\
$5f7p$ & (7/2,3/2) & 4 & 47261 & 48682 & 1421 \\
       &           & 5 & 47422 & 48611 & 1189 \\
       &           & 3 & 47471 & 48812 & 1340 \\
       &           & 2 & 49806 & 51656 & 1850 \\
\\
$6d7p$ & (3/2,3/2)$^o$ & 2 & 47680 & 48121 & 441 \\
       &            & 3 & 49981 & 50682 & 701 \\
       &            & 1 & 50993 & 51529 & 536 \\
       &            & 0 & 51745 & 52398 & 653 \\
\\
$5f2$ & $^1S$ & 0 & 51162 & 55896 & 4735 \\
\\
$6d7p$ & (5/2,3/2)$^o$ & 4 & 53052 & 53442 & 389 \\
       &            & 1 & 53939 & 54242 & 303 \\
       &            & 3 & 55552 & 56818 & 1266 \\
\\
$6d7p$ & $^o$ & 2 & 53152 & 53163 & 11 \\
\\
$7s7p$ & $^1P^o$ & 1 & 69001 & 70829 & 1828 \\
\hline
\hline
\end{tabular*}
\end{table}

\newpage

\begin{center}
Contracted Gaussian basis set for Th: $[17s17p15d14f6g5h3i]$
\end{center}
{\small\begin{verbatim}
        S                P                D                F
  5.36701700E+01   1.45964190E+02   5.28576450E+02   1.07788524E+03
  4.12847500E+01   6.64007200E+01   1.35380390E+02   3.65409080E+02
  3.17575000E+01   5.10774800E+01   4.33968800E+01   1.55254800E+02
  2.14348800E+01   3.92903700E+01   3.08752000E+01   7.31689300E+01
  1.22826900E+01   2.16264200E+01   2.79398100E+01   3.66856800E+01
  7.02645064E+00   1.07140428E+01   1.38126947E+01   1.92825954E+01
  4.31793783E+00   6.34493733E+00   8.03611544E+00   1.04209100E+01
  2.57034756E+00   3.69133896E+00   4.62685191E+00   5.54947315E+00
  1.53360871E+00   2.13493176E+00   2.59690521E+00   2.85462934E+00
  9.25765905E-01   1.21735768E+00   1.43469157E+00   1.36162382E+00
  5.32736964E-01   6.76095705E-01   7.66579414E-01   6.09628582E-01
  3.07322994E-01   3.69040410E-01   3.78374682E-01   2.52838789E-01
  1.72630009E-01   1.98168145E-01   1.80238109E-01   9.13210536E-02
  1.11785467E-01   9.86233748E-02   8.21043655E-02   5.04047100E-02
  6.20522318E-02   4.94992351E-02   3.54288527E-02
  3.32558674E-02   2.45238114E-02
  1.73630524E-02   1.20481630E-02

G
  5.48042940E+01    -0.05216008   0.10975667   0.11900634   0.23010759   0.40023312  -0.55057266
  2.14723650E+01    -0.20852245   0.44089722   0.43535446   0.48807939   0.28136182   0.29711451
  8.73159940E+00    -0.25123808   0.36669364   0.11004711  -0.60666480  -1.09734547   0.44936195
  3.97082170E+00    -0.31688226   0.03676084  -0.65337212  -0.41409547   1.14998836  -1.33634544
  1.94839610E+00    -0.32915829  -0.24662569  -0.20547070   0.89571072  -0.36526239   1.65745323
  8.09472840E-01    -0.28768056  -0.42806653   0.52049065  -0.12701185  -0.58229197  -1.36231118
  3.62200630E-01    -0.07065785  -0.12163762   0.24753720  -0.47578615   0.69911700   0.68988476
  1.25405870E-01    -0.01223940  -0.02001977   0.03267244  -0.07112408   0.14891557   0.27533913
H
  3.86539440E+01     0.04576024  -0.20059738  -0.30440567  -0.45801945  -0.74873326
  1.48410360E+01     0.09884256  -0.47026800  -0.56340601  -0.22567296   0.62180714
  7.05642780E+00     0.24622056  -0.25336193   0.20847715   0.92243234   0.20228753
  3.35510090E+00     0.38917461  -0.21780230   0.65127231  -0.40082312  -1.03230708
  1.71754750E+00     0.42826204   0.49893232  -0.31657951  -0.42458572   1.00865579
  5.09950300E-01     0.26424815   0.41994691  -0.57505676   0.65437179  -0.45595996
  2.77580970E-01    -0.06986946  -0.09564673   0.11552375  -0.04861336  -0.12373787
I
  9.60000000E+00     0.21662787   0.59188189   1.07789472
  4.80000000E+00     0.40293423   0.21708623  -0.95694527
  2.40000000E+00     0.40351893  -0.12131058  -0.09849543
  1.20000000E+00     0.24229589  -0.54814606   0.30122869
  6.00000000E-01     0.08603710  -0.17788800   0.31484677
\end{verbatim}}

\newpage

\begin{center}
Contracted Gaussian basis set for Th: $[17s17p15d14f5g4h2i]$
\end{center}
{\small\begin{verbatim}
        S                P                D                F
  5.36701700E+01   1.45964190E+02   5.28576450E+02   1.07788524E+03
  4.12847500E+01   6.64007200E+01   1.35380390E+02   3.65409080E+02
  3.17575000E+01   5.10774800E+01   4.33968800E+01   1.55254800E+02
  2.14348800E+01   3.92903700E+01   3.08752000E+01   7.31689300E+01
  1.22826900E+01   2.16264200E+01   2.79398100E+01   3.66856800E+01
  7.02645064E+00   1.07140428E+01   1.38126947E+01   1.92825954E+01
  4.31793783E+00   6.34493733E+00   8.03611544E+00   1.04209100E+01
  2.57034756E+00   3.69133896E+00   4.62685191E+00   5.54947315E+00
  1.53360871E+00   2.13493176E+00   2.59690521E+00   2.85462934E+00
  9.25765905E-01   1.21735768E+00   1.43469157E+00   1.36162382E+00
  5.32736964E-01   6.76095705E-01   7.66579414E-01   6.09628582E-01
  3.07322994E-01   3.69040410E-01   3.78374682E-01   2.52838789E-01
  1.72630009E-01   1.98168145E-01   1.80238109E-01   9.13210536E-02
  1.11785467E-01   9.86233748E-02   8.21043655E-02   5.04047100E-02
  6.20522318E-02   4.94992351E-02   3.54288527E-02
  3.32558674E-02   2.45238114E-02
  1.73630524E-02   1.20481630E-02

G
  5.48042940E+01    -0.05216008   0.10975667   0.11900634   0.23010759   0.40023312
  2.14723650E+01    -0.20852245   0.44089722   0.43535446   0.48807939   0.28136182
  8.73159940E+00    -0.25123808   0.36669364   0.11004711  -0.60666480  -1.09734547
  3.97082170E+00    -0.31688226   0.03676084  -0.65337212  -0.41409547   1.14998836
  1.94839610E+00    -0.32915829  -0.24662569  -0.20547070   0.89571072  -0.36526239
  8.09472840E-01    -0.28768056  -0.42806653   0.52049065  -0.12701185  -0.58229197
  3.62200630E-01    -0.07065785  -0.12163762   0.24753720  -0.47578615   0.69911700
  1.25405870E-01    -0.01223940  -0.02001977   0.03267244  -0.07112408   0.14891557
H
  3.86539440E+01     0.04576024  -0.20059738  -0.30440567  -0.45801945
  1.48410360E+01     0.09884256  -0.47026800  -0.56340601  -0.22567296
  7.05642780E+00     0.24622056  -0.25336193   0.20847715   0.92243234
  3.35510090E+00     0.38917461  -0.21780230   0.65127231  -0.40082312
  1.71754750E+00     0.42826204   0.49893232  -0.31657951  -0.42458572
  5.09950300E-01     0.26424815   0.41994691  -0.57505676   0.65437179
  2.77580970E-01    -0.06986946  -0.09564673   0.11552375  -0.04861336
I
  9.60000000E+00     0.21662787   0.59188189
  4.80000000E+00     0.40293423   0.21708623
  2.40000000E+00     0.40351893  -0.12131058
  1.20000000E+00     0.24229589  -0.54814606
  6.00000000E-01     0.08603710  -0.17788800
\end{verbatim}}

\newpage

\begin{center}
Contracted Gaussian basis set for Th: $[17s17p15d14f4g3h1i]$
\end{center}
{\small\begin{verbatim}
        S                P                D                F
  5.36701700E+01   1.45964190E+02   5.28576450E+02   1.07788524E+03
  4.12847500E+01   6.64007200E+01   1.35380390E+02   3.65409080E+02
  3.17575000E+01   5.10774800E+01   4.33968800E+01   1.55254800E+02
  2.14348800E+01   3.92903700E+01   3.08752000E+01   7.31689300E+01
  1.22826900E+01   2.16264200E+01   2.79398100E+01   3.66856800E+01
  7.02645064E+00   1.07140428E+01   1.38126947E+01   1.92825954E+01
  4.31793783E+00   6.34493733E+00   8.03611544E+00   1.04209100E+01
  2.57034756E+00   3.69133896E+00   4.62685191E+00   5.54947315E+00
  1.53360871E+00   2.13493176E+00   2.59690521E+00   2.85462934E+00
  9.25765905E-01   1.21735768E+00   1.43469157E+00   1.36162382E+00
  5.32736964E-01   6.76095705E-01   7.66579414E-01   6.09628582E-01
  3.07322994E-01   3.69040410E-01   3.78374682E-01   2.52838789E-01
  1.72630009E-01   1.98168145E-01   1.80238109E-01   9.13210536E-02
  1.11785467E-01   9.86233748E-02   8.21043655E-02   5.04047100E-02
  6.20522318E-02   4.94992351E-02   3.54288527E-02
  3.32558674E-02   2.45238114E-02
  1.73630524E-02   1.20481630E-02

G
  5.48042940E+01    -0.05216008   0.10975667   0.11900634   0.23010759
  2.14723650E+01    -0.20852245   0.44089722   0.43535446   0.48807939
  8.73159940E+00    -0.25123808   0.36669364   0.11004711  -0.60666480
  3.97082170E+00    -0.31688226   0.03676084  -0.65337212  -0.41409547
  1.94839610E+00    -0.32915829  -0.24662569  -0.20547070   0.89571072
  8.09472840E-01    -0.28768056  -0.42806653   0.52049065  -0.12701185
  3.62200630E-01    -0.07065785  -0.12163762   0.24753720  -0.47578615
  1.25405870E-01    -0.01223940  -0.02001977   0.03267244  -0.07112408
H
  3.86539440E+01     0.04576024  -0.20059738  -0.30440567
  1.48410360E+01     0.09884256  -0.47026800  -0.56340601
  7.05642780E+00     0.24622056  -0.25336193   0.20847715
  3.35510090E+00     0.38917461  -0.21780230   0.65127231
  1.71754750E+00     0.42826204   0.49893232  -0.31657951
  5.09950300E-01     0.26424815   0.41994691  -0.57505676
  2.77580970E-01    -0.06986946  -0.09564673   0.11552375
I
  9.60000000E+00     0.21662787
  4.80000000E+00     0.40293423
  2.40000000E+00     0.40351893
  1.20000000E+00     0.24229589
  6.00000000E-01     0.08603710
\end{verbatim}}

\newpage

\begin{center}
Contracted Gaussian basis set for Th: $[9s10p9d8f6g5h3i]$
\end{center}
{\small\begin{verbatim}
S
  5.36701700E+01     0.31235107   0.36263714   0.26438208  -0.11812526   0.68018455
  4.12847500E+01    -1.42427639  -0.38223740  -0.40944780   0.19517531  -1.29886872
  3.17575000E+01     1.92723004  -0.69364101  -0.21336713   0.07128691  -0.21317826
  2.14348800E+01     0.11557486   0.13303689   0.07730284  -0.03548529   0.54810785
  1.22826900E+01    -1.46996032   0.94364992   0.49779738  -0.21212683   1.15949048
  7.02645064E+00     0.67786446   0.83281455   0.67204692  -0.31091590   0.85362234
  4.31793783E+00    -1.54187939  -0.09355498  -0.30117401   0.17201400  -1.89510193
  2.57034756E+00     1.34679483  -0.98265388  -0.82111893   0.42174203  -1.55498936
  1.53360871E+00    -1.45751235  -0.19887999  -0.71145926   0.43098311   0.35100783
  9.25765905E-01     1.16546053  -0.29569302   0.20474776  -0.25718535   3.13122148
  5.32736964E-01    -0.89755444   0.10896496   0.70924478  -0.61759534  -0.05244383
  3.07322994E-01     0.69329403  -0.19013689   0.54206347  -0.64187672  -2.03165477
  1.72630009E-01    -0.58084686   0.12208461   0.09190060  -0.07899636  -0.72812696
  1.11785467E-01     0.39212548  -0.09083827   0.00005679   0.87826771   0.73482833
  6.20522318E-02    -0.14982819   0.03585468   0.00197485   0.62835283   0.51777336
  3.32558674E-02     0.05003330  -0.01229848  -0.00030560   0.01628351   0.02590714
  1.73630524E-02    -0.01019895   0.00254537   0.00008232   0.00235392   0.00037172
S
  5.36701700E+01    -1.67439834  -2.86693990  -4.45697539  -9.13065395
  4.12847500E+01     3.53037713   7.02616334  12.64413558  29.94981655
  3.17575000E+01     0.46331671  -0.66722788  -4.90883012 -22.52266585
  2.14348800E+01    -3.07314605  -7.34507349 -12.04094449 -13.55907675
  1.22826900E+01    -1.81691919   1.26282487  13.28429206  37.63295703
  7.02645064E+00     1.52438667   7.85165087   0.64483700 -41.66446404
  4.31793783E+00     4.70359414  -3.68462837 -13.50341941  20.95741453
  2.57034756E+00    -3.20225018  -9.37515263   7.71438030   8.53783105
  1.53360871E+00    -4.15718825  11.21410520  11.29294125 -19.72106623
  9.25765905E-01     3.07321173  -0.64060204 -20.74256018   8.62667179
  5.32736964E-01     3.16715318  -5.36492768  12.65155732   8.73886496
  3.07322994E-01    -2.84592411   1.09973570  -0.99786643 -15.08069268
  1.72630009E-01    -0.73408993   2.73426419  -1.35229908   9.64538748
  1.11785467E-01     0.27713900  -0.72165769  -2.18815103  -2.07731224
  6.20522318E-02     0.70266592  -0.68861067   1.93042988   0.15676516
  3.32558674E-02    -0.00658111  -0.12950112   0.02294316  -0.89551960
  1.73630524E-02     0.00928795   0.01056190   0.03196954   0.12585266
P
  1.45964190E+02     0.00530765   0.00525740   0.00309793   0.00407496  -0.03540133
  6.64007200E+01     0.21872146   0.13631905   0.04976924  -0.06009810   0.67013584
  5.10774800E+01    -0.68931136  -0.50295087  -0.21858156   0.15262427  -1.82081815
  3.92903700E+01     0.70259096   0.70372929   0.39298057  -0.02418076   1.02308919
  2.16264200E+01    -0.21438074  -0.35908861  -0.25728148  -0.16925999   0.96925663
  1.07140428E+01    -0.29587998  -0.66539070  -0.51382085  -0.19016172   0.31207454
  6.34493733E+00    -0.17626810  -0.21214491  -0.10633060   0.10619757  -1.04797653
  3.69133896E+00    -0.25420814   0.20897237   0.34453633   0.22213180  -0.59661347
  2.13493176E+00    -0.26294866   0.36717237   0.58464545   0.21427410   0.26223177
  1.21735768E+00    -0.11423645   0.16949879   0.15368325  -0.10248942   0.91853872
  6.76095705E-01    -0.01855420   0.16350754  -0.33495401  -0.32633923   0.24266073
  3.69040410E-01    -0.00298398   0.17460870  -0.53586485  -0.31524275  -0.52815531
  1.98168145E-01    -0.00099576   0.08709730  -0.25592810  -0.07526693  -0.60036035
  9.86233748E-02     0.00010906   0.00433478  -0.02101408   0.66658637   0.34089390
  4.94992351E-02    -0.00008199   0.00082115   0.00072971   0.48906441   0.32101387
  2.45238114E-02     0.00003573  -0.00032629  -0.00047471  -0.00740507  -0.00265507
  1.20481630E-02    -0.00001017   0.00009647   0.00010556   0.00565506   0.00327090
\end{verbatim}}

\newpage

{\small\begin{verbatim}
P
  1.45964190E+02    -0.07010951   0.11145797   0.13208927   0.13536568  -0.02288599
  6.64007200E+01     1.48522236  -2.52847590  -3.27423460  -4.27820944   2.98494455
  5.10774800E+01    -3.77660123   6.13122434   8.65890415  14.23948467 -15.68307065
  3.92903700E+01     2.40668034  -4.03809253  -6.76542113 -14.50257499  21.27462828
  2.16264200E+01     1.25719704  -1.32715385  -0.26537825   4.70880026 -14.23092917
  1.07140428E+01    -0.93788462   2.74199459   4.43696439   3.37479514   9.87354080
  6.34493733E+00    -1.25512433   0.31402823  -3.53913396  -9.13130071  -2.21042107
  3.69133896E+00     0.69013232  -3.48701616  -2.71746748   7.43197666  -8.32942791
  2.13493176E+00     1.56386564   0.98600075   6.52603270   0.10792641  13.96020127
  1.21735768E+00    -0.59682925   3.08907471  -3.55530905  -6.45680360 -12.28446148
  6.76095705E-01    -1.54107709  -2.44598884  -1.92594140   7.18257162   6.00379278
  3.69040410E-01     0.41776292  -0.92466919   3.73251207  -3.38349650   0.31317671
  1.98168145E-01     1.19754917   1.60833446  -1.22933242  -0.83701228  -3.39771829
  9.86233748E-02    -0.28202923   0.19430444  -1.36462683   2.31985617   3.00549943
  4.94992351E-02    -0.46963884  -0.70422600   1.09793506  -1.08566747  -1.03001238
  2.45238114E-02     0.00270849  -0.01653011   0.10568627  -0.19996210  -0.21195079
  1.20481630E-02    -0.00528011  -0.00705279   0.00194501   0.00769251  -0.00013002
D
  5.28576450E+02     0.00056610  -0.00033618  -0.00012061   0.00065414  -0.00152210
  1.35380390E+02     0.00804706  -0.00359210  -0.00123909   0.00496550  -0.00808237
  4.33968800E+01    -0.10272249   0.00893315   0.00243988  -0.07629249   0.25289006
  3.08752000E+01     0.71863038  -0.27671750  -0.09670638   0.71644532  -1.92383097
  2.79398100E+01    -0.79927936   0.38788168   0.14082683  -0.94205956   2.44395855
  1.38126947E+01    -0.41788407   0.27890012   0.09599948  -0.23341652   0.10046912
  8.03611544E+00    -0.32277375   0.16262090   0.04846383   0.02380230  -0.63074613
  4.62685191E+00    -0.14213270  -0.16211722  -0.09191526   0.49969141  -0.67451923
  2.59690521E+00    -0.03923363  -0.44310225  -0.20799912   0.42756263   0.49064726
  1.43469157E+00    -0.02332129  -0.39009976  -0.16273560  -0.20624138   0.95925792
  7.66579414E-01    -0.00791208  -0.15994484   0.12903953  -0.78392624  -0.63404762
  3.78374682E-01    -0.00141901  -0.02028650   0.41170550  -0.13807411  -0.65285894
  1.80238109E-01     0.00018550   0.00024080   0.47673105   0.52217198   0.47625071
  8.21043655E-02    -0.00008170  -0.00024139   0.16859589   0.21536222   0.25600883
  3.54288527E-02     0.00002271   0.00004912  -0.00189440   0.00026765   0.00307811
D
  5.28576450E+02    -0.00199163   0.00294419  -0.00500586  -0.00880390
  1.35380390E+02    -0.00648153   0.00104152   0.01708301   0.06607903
  4.33968800E+01     0.32400998  -0.06891424  -0.60112456  -1.71896872
  3.08752000E+01    -2.20451061  -0.57917875   8.40275543  22.89765433
  2.79398100E+01     2.93999810  -0.49106249  -7.31565088 -23.16100558
  1.38126947E+01    -0.51573270   1.59740272  -2.24099441   1.44006755
  8.03611544E+00    -1.26293676   0.48470649   3.27877777   4.07590145
  4.62685191E+00     0.71370037  -3.01712350  -0.35226550  -7.10900711
  2.59690521E+00     1.55528878   2.22448999  -3.32469633   6.15838487
  1.43469157E+00    -1.65059666   0.73660737   4.44534731  -2.90258119
  7.66579414E-01    -0.34619940  -2.28243693  -2.79777474  -0.25005207
  3.78374682E-01     1.26487320   1.43418304   0.41686773   1.89935458
  1.80238109E-01    -0.36288291   0.11271919   0.96477752  -1.99734624
  8.21043655E-02    -0.37686010  -0.58299714  -0.81575839   0.98325096
  3.54288527E-02    -0.01009148  -0.02810431  -0.05640283   0.08403634
\end{verbatim}
}

\newpage

{\small\begin{verbatim}
F
  1.07788524E+03     0.00066139  -0.00019180   0.00062963  -0.00123311   0.00168667
  3.65409080E+02     0.00611650  -0.00178165   0.00516386  -0.00914729   0.01105597
  1.55254800E+02     0.03221673  -0.00924938   0.02355697  -0.04020323   0.05116414
  7.31689300E+01     0.10757267  -0.03129252   0.08037328  -0.13224258   0.15072076
  3.66856800E+01     0.23886728  -0.06756767   0.15521256  -0.25005590   0.32161297
  1.92825954E+01     0.32998174  -0.08901212   0.22808388  -0.27860704   0.14563047
  1.04209100E+01     0.32230710  -0.06439913   0.04886440   0.34723392  -0.70315325
  5.54947315E+00     0.19639736   0.04495256  -0.27419029   0.52302944  -0.24474715
  2.85462934E+00     0.05633795   0.22284324  -0.58380604  -0.06922958   0.96931251
  1.36162382E+00     0.00420808   0.36510976  -0.07657812  -0.69021627  -0.26060962
  6.09628582E-01    -0.00000225   0.37908671   0.41344367   0.02083569  -0.74710139
  2.52838789E-01     0.00001167   0.25199346   0.29412640   0.50281350   0.62467306
  9.13210536E-02    -0.00000244   0.06323981   0.07431200   0.15946321   0.26560814
  5.04047100E-02     0.00000115  -0.00870342  -0.00969336  -0.01783575  -0.02862745
F
  1.07788524E+03     0.00252560   0.00468749   0.00668247
  3.65409080E+02     0.01321445   0.01242463   0.00828996
  1.55254800E+02     0.07446308   0.11816457   0.14068472
  7.31689300E+01     0.16488129   0.07508034  -0.08389073
  3.66856800E+01     0.52195480   0.93495093   1.12952307
  1.92825954E+01    -0.39905538  -1.89372408  -2.96939939
  1.04209100E+01    -1.02578092   0.85578325   3.78975824
  5.54947315E+00     1.35136455   1.19083340  -2.89085267
  2.85462934E+00    -0.01548639  -2.16452138   1.21400251
  1.36162382E+00    -1.17681923   1.74061765   0.17586282
  6.09628582E-01     1.16486834  -0.72541869  -0.97375909
  2.52838789E-01    -0.33844950  -0.20477002   1.27689416
  9.13210536E-02    -0.35504014   0.62680445  -1.16793384
  5.04047100E-02     0.05272103  -0.12749049   0.24816875
G
  5.48042940E+01    -0.05216008   0.10975667   0.11900634   0.23010759   0.40023312  -0.55057266
  2.14723650E+01    -0.20852245   0.44089722   0.43535446   0.48807939   0.28136182   0.29711451
  8.73159940E+00    -0.25123808   0.36669364   0.11004711  -0.60666480  -1.09734547   0.44936195
  3.97082170E+00    -0.31688226   0.03676084  -0.65337212  -0.41409547   1.14998836  -1.33634544
  1.94839610E+00    -0.32915829  -0.24662569  -0.20547070   0.89571072  -0.36526239   1.65745323
  8.09472840E-01    -0.28768056  -0.42806653   0.52049065  -0.12701185  -0.58229197  -1.36231118
  3.62200630E-01    -0.07065785  -0.12163762   0.24753720  -0.47578615   0.69911700   0.68988476
  1.25405870E-01    -0.01223940  -0.02001977   0.03267244  -0.07112408   0.14891557   0.27533913
H
  3.86539440E+01     0.04576024  -0.20059738  -0.30440567  -0.45801945  -0.74873326
  1.48410360E+01     0.09884256  -0.47026800  -0.56340601  -0.22567296   0.62180714
  7.05642780E+00     0.24622056  -0.25336193   0.20847715   0.92243234   0.20228753
  3.35510090E+00     0.38917461  -0.21780230   0.65127231  -0.40082312  -1.03230708
  1.71754750E+00     0.42826204   0.49893232  -0.31657951  -0.42458572   1.00865579
  5.09950300E-01     0.26424815   0.41994691  -0.57505676   0.65437179  -0.45595996
  2.77580970E-01    -0.06986946  -0.09564673   0.11552375  -0.04861336  -0.12373787
I
  9.60000000E+00     0.21662787   0.59188189   1.07789472
  4.80000000E+00     0.40293423   0.21708623  -0.95694527
  2.40000000E+00     0.40351893  -0.12131058  -0.09849543
  1.20000000E+00     0.24229589  -0.54814606   0.30122869
  6.00000000E-01     0.08603710  -0.17788800   0.31484677
\end{verbatim}
}

\newpage

\begin{center}
Contracted Gaussian basis set for Th: $[Th8s9p8d7f5g4h2i]$
\end{center}
{\small\begin{verbatim}
S
  5.36701700E+01     0.31235107   0.36263714   0.26438208  -0.11812526   0.68018455
  4.12847500E+01    -1.42427639  -0.38223740  -0.40944780   0.19517531  -1.29886872
  3.17575000E+01     1.92723004  -0.69364101  -0.21336713   0.07128691  -0.21317826
  2.14348800E+01     0.11557486   0.13303689   0.07730284  -0.03548529   0.54810785
  1.22826900E+01    -1.46996032   0.94364992   0.49779738  -0.21212683   1.15949048
  7.02645064E+00     0.67786446   0.83281455   0.67204692  -0.31091590   0.85362234
  4.31793783E+00    -1.54187939  -0.09355498  -0.30117401   0.17201400  -1.89510193
  2.57034756E+00     1.34679483  -0.98265388  -0.82111893   0.42174203  -1.55498936
  1.53360871E+00    -1.45751235  -0.19887999  -0.71145926   0.43098311   0.35100783
  9.25765905E-01     1.16546053  -0.29569302   0.20474776  -0.25718535   3.13122148
  5.32736964E-01    -0.89755444   0.10896496   0.70924478  -0.61759534  -0.05244383
  3.07322994E-01     0.69329403  -0.19013689   0.54206347  -0.64187672  -2.03165477
  1.72630009E-01    -0.58084686   0.12208461   0.09190060  -0.07899636  -0.72812696
  1.11785467E-01     0.39212548  -0.09083827   0.00005679   0.87826771   0.73482833
  6.20522318E-02    -0.14982819   0.03585468   0.00197485   0.62835283   0.51777336
  3.32558674E-02     0.05003330  -0.01229848  -0.00030560   0.01628351   0.02590714
  1.73630524E-02    -0.01019895   0.00254537   0.00008232   0.00235392   0.00037172
S
  5.36701700E+01    -1.67439834  -2.86693990  -4.45697539
  4.12847500E+01     3.53037713   7.02616334  12.64413558
  3.17575000E+01     0.46331671  -0.66722788  -4.90883012
  2.14348800E+01    -3.07314605  -7.34507349 -12.04094449
  1.22826900E+01    -1.81691919   1.26282487  13.28429206
  7.02645064E+00     1.52438667   7.85165087   0.64483700
  4.31793783E+00     4.70359414  -3.68462837 -13.50341941
  2.57034756E+00    -3.20225018  -9.37515263   7.71438030
  1.53360871E+00    -4.15718825  11.21410520  11.29294125
  9.25765905E-01     3.07321173  -0.64060204 -20.74256018
  5.32736964E-01     3.16715318  -5.36492768  12.65155732
  3.07322994E-01    -2.84592411   1.09973570  -0.99786643
  1.72630009E-01    -0.73408993   2.73426419  -1.35229908
  1.11785467E-01     0.27713900  -0.72165769  -2.18815103
  6.20522318E-02     0.70266592  -0.68861067   1.93042988
  3.32558674E-02    -0.00658111  -0.12950112   0.02294316
  1.73630524E-02     0.00928795   0.01056190   0.03196954
P
  1.45964190E+02     0.00530765   0.00525740   0.00309793   0.00407496  -0.03540133
  6.64007200E+01     0.21872146   0.13631905   0.04976924  -0.06009810   0.67013584
  5.10774800E+01    -0.68931136  -0.50295087  -0.21858156   0.15262427  -1.82081815
  3.92903700E+01     0.70259096   0.70372929   0.39298057  -0.02418076   1.02308919
  2.16264200E+01    -0.21438074  -0.35908861  -0.25728148  -0.16925999   0.96925663
  1.07140428E+01    -0.29587998  -0.66539070  -0.51382085  -0.19016172   0.31207454
  6.34493733E+00    -0.17626810  -0.21214491  -0.10633060   0.10619757  -1.04797653
  3.69133896E+00    -0.25420814   0.20897237   0.34453633   0.22213180  -0.59661347
  2.13493176E+00    -0.26294866   0.36717237   0.58464545   0.21427410   0.26223177
  1.21735768E+00    -0.11423645   0.16949879   0.15368325  -0.10248942   0.91853872
  6.76095705E-01    -0.01855420   0.16350754  -0.33495401  -0.32633923   0.24266073
  3.69040410E-01    -0.00298398   0.17460870  -0.53586485  -0.31524275  -0.52815531
  1.98168145E-01    -0.00099576   0.08709730  -0.25592810  -0.07526693  -0.60036035
  9.86233748E-02     0.00010906   0.00433478  -0.02101408   0.66658637   0.34089390
  4.94992351E-02    -0.00008199   0.00082115   0.00072971   0.48906441   0.32101387
  2.45238114E-02     0.00003573  -0.00032629  -0.00047471  -0.00740507  -0.00265507
  1.20481630E-02    -0.00001017   0.00009647   0.00010556   0.00565506   0.00327090
\end{verbatim}
}

\newpage

{\small\begin{verbatim}
P
  1.45964190E+02    -0.07010951   0.11145797   0.13208927   0.13536568
  6.64007200E+01     1.48522236  -2.52847590  -3.27423460  -4.27820944
  5.10774800E+01    -3.77660123   6.13122434   8.65890415  14.23948467
  3.92903700E+01     2.40668034  -4.03809253  -6.76542113 -14.50257499
  2.16264200E+01     1.25719704  -1.32715385  -0.26537825   4.70880026
  1.07140428E+01    -0.93788462   2.74199459   4.43696439   3.37479514
  6.34493733E+00    -1.25512433   0.31402823  -3.53913396  -9.13130071
  3.69133896E+00     0.69013232  -3.48701616  -2.71746748   7.43197666
  2.13493176E+00     1.56386564   0.98600075   6.52603270   0.10792641
  1.21735768E+00    -0.59682925   3.08907471  -3.55530905  -6.45680360
  6.76095705E-01    -1.54107709  -2.44598884  -1.92594140   7.18257162
  3.69040410E-01     0.41776292  -0.92466919   3.73251207  -3.38349650
  1.98168145E-01     1.19754917   1.60833446  -1.22933242  -0.83701228
  9.86233748E-02    -0.28202923   0.19430444  -1.36462683   2.31985617
  4.94992351E-02    -0.46963884  -0.70422600   1.09793506  -1.08566747
  2.45238114E-02     0.00270849  -0.01653011   0.10568627  -0.19996210
  1.20481630E-02    -0.00528011  -0.00705279   0.00194501   0.00769251
D
  5.28576450E+02     0.00056610  -0.00033618  -0.00012061   0.00065414  -0.00152210
  1.35380390E+02     0.00804706  -0.00359210  -0.00123909   0.00496550  -0.00808237
  4.33968800E+01    -0.10272249   0.00893315   0.00243988  -0.07629249   0.25289006
  3.08752000E+01     0.71863038  -0.27671750  -0.09670638   0.71644532  -1.92383097
  2.79398100E+01    -0.79927936   0.38788168   0.14082683  -0.94205956   2.44395855
  1.38126947E+01    -0.41788407   0.27890012   0.09599948  -0.23341652   0.10046912
  8.03611544E+00    -0.32277375   0.16262090   0.04846383   0.02380230  -0.63074613
  4.62685191E+00    -0.14213270  -0.16211722  -0.09191526   0.49969141  -0.67451923
  2.59690521E+00    -0.03923363  -0.44310225  -0.20799912   0.42756263   0.49064726
  1.43469157E+00    -0.02332129  -0.39009976  -0.16273560  -0.20624138   0.95925792
  7.66579414E-01    -0.00791208  -0.15994484   0.12903953  -0.78392624  -0.63404762
  3.78374682E-01    -0.00141901  -0.02028650   0.41170550  -0.13807411  -0.65285894
  1.80238109E-01     0.00018550   0.00024080   0.47673105   0.52217198   0.47625071
  8.21043655E-02    -0.00008170  -0.00024139   0.16859589   0.21536222   0.25600883
  3.54288527E-02     0.00002271   0.00004912  -0.00189440   0.00026765   0.00307811
D
  5.28576450E+02    -0.00199163   0.00294419  -0.00500586
  1.35380390E+02    -0.00648153   0.00104152   0.01708301
  4.33968800E+01     0.32400998  -0.06891424  -0.60112456
  3.08752000E+01    -2.20451061  -0.57917875   8.40275543
  2.79398100E+01     2.93999810  -0.49106249  -7.31565088
  1.38126947E+01    -0.51573270   1.59740272  -2.24099441
  8.03611544E+00    -1.26293676   0.48470649   3.27877777
  4.62685191E+00     0.71370037  -3.01712350  -0.35226550
  2.59690521E+00     1.55528878   2.22448999  -3.32469633
  1.43469157E+00    -1.65059666   0.73660737   4.44534731
  7.66579414E-01    -0.34619940  -2.28243693  -2.79777474
  3.78374682E-01     1.26487320   1.43418304   0.41686773
  1.80238109E-01    -0.36288291   0.11271919   0.96477752
  8.21043655E-02    -0.37686010  -0.58299714  -0.81575839
  3.54288527E-02    -0.01009148  -0.02810431  -0.05640283
\end{verbatim}
}

\newpage

{\small\begin{verbatim}
F
  1.07788524E+03     0.00066139  -0.00019180   0.00062963  -0.00123311   0.00168667   0.00252560   0.00468749
  3.65409080E+02     0.00611650  -0.00178165   0.00516386  -0.00914729   0.01105597   0.01321445   0.01242463
  1.55254800E+02     0.03221673  -0.00924938   0.02355697  -0.04020323   0.05116414   0.07446308   0.11816457
  7.31689300E+01     0.10757267  -0.03129252   0.08037328  -0.13224258   0.15072076   0.16488129   0.07508034
  3.66856800E+01     0.23886728  -0.06756767   0.15521256  -0.25005590   0.32161297   0.52195480   0.93495093
  1.92825954E+01     0.32998174  -0.08901212   0.22808388  -0.27860704   0.14563047  -0.39905538  -1.89372408
  1.04209100E+01     0.32230710  -0.06439913   0.04886440   0.34723392  -0.70315325  -1.02578092   0.85578325
  5.54947315E+00     0.19639736   0.04495256  -0.27419029   0.52302944  -0.24474715   1.35136455   1.19083340
  2.85462934E+00     0.05633795   0.22284324  -0.58380604  -0.06922958   0.96931251  -0.01548639  -2.16452138
  1.36162382E+00     0.00420808   0.36510976  -0.07657812  -0.69021627  -0.26060962  -1.17681923   1.74061765
  6.09628582E-01    -0.00000225   0.37908671   0.41344367   0.02083569  -0.74710139   1.16486834  -0.72541869
  2.52838789E-01     0.00001167   0.25199346   0.29412640   0.50281350   0.62467306  -0.33844950  -0.20477002
  9.13210536E-02    -0.00000244   0.06323981   0.07431200   0.15946321   0.26560814  -0.35504014   0.62680445
  5.04047100E-02     0.00000115  -0.00870342  -0.00969336  -0.01783575  -0.02862745   0.05272103  -0.12749049
G
  5.48042940E+01    -0.05216008   0.10975667   0.11900634   0.23010759   0.40023312
  2.14723650E+01    -0.20852245   0.44089722   0.43535446   0.48807939   0.28136182
  8.73159940E+00    -0.25123808   0.36669364   0.11004711  -0.60666480  -1.09734547
  3.97082170E+00    -0.31688226   0.03676084  -0.65337212  -0.41409547   1.14998836
  1.94839610E+00    -0.32915829  -0.24662569  -0.20547070   0.89571072  -0.36526239
  8.09472840E-01    -0.28768056  -0.42806653   0.52049065  -0.12701185  -0.58229197
  3.62200630E-01    -0.07065785  -0.12163762   0.24753720  -0.47578615   0.69911700
  1.25405870E-01    -0.01223940  -0.02001977   0.03267244  -0.07112408   0.14891557
H
  3.86539440E+01     0.04576024  -0.20059738  -0.30440567  -0.45801945
  1.48410360E+01     0.09884256  -0.47026800  -0.56340601  -0.22567296
  7.05642780E+00     0.24622056  -0.25336193   0.20847715   0.92243234
  3.35510090E+00     0.38917461  -0.21780230   0.65127231  -0.40082312
  1.71754750E+00     0.42826204   0.49893232  -0.31657951  -0.42458572
  5.09950300E-01     0.26424815   0.41994691  -0.57505676   0.65437179
  2.77580970E-01    -0.06986946  -0.09564673   0.11552375  -0.04861336
I
  9.60000000E+00     0.21662787   0.59188189
  4.80000000E+00     0.40293423   0.21708623
  2.40000000E+00     0.40351893  -0.12131058
  1.20000000E+00     0.24229589  -0.54814606
  6.00000000E-01     0.08603710  -0.17788800
\end{verbatim}
}

\newpage

\begin{center}
Contracted Gaussian basis set for Th: $[8s8p7d6f4g3h1i]$
\end{center}
{\small\begin{verbatim}
S
  5.36701700E+01     0.31235107   0.36263714   0.26438208  -0.11812526   0.68018455
  4.12847500E+01    -1.42427639  -0.38223740  -0.40944780   0.19517531  -1.29886872
  3.17575000E+01     1.92723004  -0.69364101  -0.21336713   0.07128691  -0.21317826
  2.14348800E+01     0.11557486   0.13303689   0.07730284  -0.03548529   0.54810785
  1.22826900E+01    -1.46996032   0.94364992   0.49779738  -0.21212683   1.15949048
  7.02645064E+00     0.67786446   0.83281455   0.67204692  -0.31091590   0.85362234
  4.31793783E+00    -1.54187939  -0.09355498  -0.30117401   0.17201400  -1.89510193
  2.57034756E+00     1.34679483  -0.98265388  -0.82111893   0.42174203  -1.55498936
  1.53360871E+00    -1.45751235  -0.19887999  -0.71145926   0.43098311   0.35100783
  9.25765905E-01     1.16546053  -0.29569302   0.20474776  -0.25718535   3.13122148
  5.32736964E-01    -0.89755444   0.10896496   0.70924478  -0.61759534  -0.05244383
  3.07322994E-01     0.69329403  -0.19013689   0.54206347  -0.64187672  -2.03165477
  1.72630009E-01    -0.58084686   0.12208461   0.09190060  -0.07899636  -0.72812696
  1.11785467E-01     0.39212548  -0.09083827   0.00005679   0.87826771   0.73482833
  6.20522318E-02    -0.14982819   0.03585468   0.00197485   0.62835283   0.51777336
  3.32558674E-02     0.05003330  -0.01229848  -0.00030560   0.01628351   0.02590714
  1.73630524E-02    -0.01019895   0.00254537   0.00008232   0.00235392   0.00037172
S
  5.36701700E+01    -1.67439834  -2.86693990  -4.45697539
  4.12847500E+01     3.53037713   7.02616334  12.64413558
  3.17575000E+01     0.46331671  -0.66722788  -4.90883012
  2.14348800E+01    -3.07314605  -7.34507349 -12.04094449
  1.22826900E+01    -1.81691919   1.26282487  13.28429206
  7.02645064E+00     1.52438667   7.85165087   0.64483700
  4.31793783E+00     4.70359414  -3.68462837 -13.50341941
  2.57034756E+00    -3.20225018  -9.37515263   7.71438030
  1.53360871E+00    -4.15718825  11.21410520  11.29294125
  9.25765905E-01     3.07321173  -0.64060204 -20.74256018
  5.32736964E-01     3.16715318  -5.36492768  12.65155732
  3.07322994E-01    -2.84592411   1.09973570  -0.99786643
  1.72630009E-01    -0.73408993   2.73426419  -1.35229908
  1.11785467E-01     0.27713900  -0.72165769  -2.18815103
  6.20522318E-02     0.70266592  -0.68861067   1.93042988
  3.32558674E-02    -0.00658111  -0.12950112   0.02294316
  1.73630524E-02     0.00928795   0.01056190   0.03196954
P
  1.45964190E+02     0.00530765   0.00525740   0.00309793   0.00407496  -0.03540133
  6.64007200E+01     0.21872146   0.13631905   0.04976924  -0.06009810   0.67013584
  5.10774800E+01    -0.68931136  -0.50295087  -0.21858156   0.15262427  -1.82081815
  3.92903700E+01     0.70259096   0.70372929   0.39298057  -0.02418076   1.02308919
  2.16264200E+01    -0.21438074  -0.35908861  -0.25728148  -0.16925999   0.96925663
  1.07140428E+01    -0.29587998  -0.66539070  -0.51382085  -0.19016172   0.31207454
  6.34493733E+00    -0.17626810  -0.21214491  -0.10633060   0.10619757  -1.04797653
  3.69133896E+00    -0.25420814   0.20897237   0.34453633   0.22213180  -0.59661347
  2.13493176E+00    -0.26294866   0.36717237   0.58464545   0.21427410   0.26223177
  1.21735768E+00    -0.11423645   0.16949879   0.15368325  -0.10248942   0.91853872
  6.76095705E-01    -0.01855420   0.16350754  -0.33495401  -0.32633923   0.24266073
  3.69040410E-01    -0.00298398   0.17460870  -0.53586485  -0.31524275  -0.52815531
  1.98168145E-01    -0.00099576   0.08709730  -0.25592810  -0.07526693  -0.60036035
  9.86233748E-02     0.00010906   0.00433478  -0.02101408   0.66658637   0.34089390
  4.94992351E-02    -0.00008199   0.00082115   0.00072971   0.48906441   0.32101387
  2.45238114E-02     0.00003573  -0.00032629  -0.00047471  -0.00740507  -0.00265507
  1.20481630E-02    -0.00001017   0.00009647   0.00010556   0.00565506   0.00327090
\end{verbatim}
}

\newpage

{\small\begin{verbatim}
P
  1.45964190E+02    -0.07010951   0.11145797   0.13208927
  6.64007200E+01     1.48522236  -2.52847590  -3.27423460
  5.10774800E+01    -3.77660123   6.13122434   8.65890415
  3.92903700E+01     2.40668034  -4.03809253  -6.76542113
  2.16264200E+01     1.25719704  -1.32715385  -0.26537825
  1.07140428E+01    -0.93788462   2.74199459   4.43696439
  6.34493733E+00    -1.25512433   0.31402823  -3.53913396
  3.69133896E+00     0.69013232  -3.48701616  -2.71746748
  2.13493176E+00     1.56386564   0.98600075   6.52603270
  1.21735768E+00    -0.59682925   3.08907471  -3.55530905
  6.76095705E-01    -1.54107709  -2.44598884  -1.92594140
  3.69040410E-01     0.41776292  -0.92466919   3.73251207
  1.98168145E-01     1.19754917   1.60833446  -1.22933242
  9.86233748E-02    -0.28202923   0.19430444  -1.36462683
  4.94992351E-02    -0.46963884  -0.70422600   1.09793506
  2.45238114E-02     0.00270849  -0.01653011   0.10568627
  1.20481630E-02    -0.00528011  -0.00705279   0.00194501
D
  5.28576450E+02     0.00056610  -0.00033618  -0.00012061   0.00065414  -0.00152210  -0.00199163   0.00294419
  1.35380390E+02     0.00804706  -0.00359210  -0.00123909   0.00496550  -0.00808237  -0.00648153   0.00104152
  4.33968800E+01    -0.10272249   0.00893315   0.00243988  -0.07629249   0.25289006   0.32400998  -0.06891424
  3.08752000E+01     0.71863038  -0.27671750  -0.09670638   0.71644532  -1.92383097  -2.20451061  -0.57917875
  2.79398100E+01    -0.79927936   0.38788168   0.14082683  -0.94205956   2.44395855   2.93999810  -0.49106249
  1.38126947E+01    -0.41788407   0.27890012   0.09599948  -0.23341652   0.10046912  -0.51573270   1.59740272
  8.03611544E+00    -0.32277375   0.16262090   0.04846383   0.02380230  -0.63074613  -1.26293676   0.48470649
  4.62685191E+00    -0.14213270  -0.16211722  -0.09191526   0.49969141  -0.67451923   0.71370037  -3.01712350
  2.59690521E+00    -0.03923363  -0.44310225  -0.20799912   0.42756263   0.49064726   1.55528878   2.22448999
  1.43469157E+00    -0.02332129  -0.39009976  -0.16273560  -0.20624138   0.95925792  -1.65059666   0.73660737
  7.66579414E-01    -0.00791208  -0.15994484   0.12903953  -0.78392624  -0.63404762  -0.34619940  -2.28243693
  3.78374682E-01    -0.00141901  -0.02028650   0.41170550  -0.13807411  -0.65285894   1.26487320   1.43418304
  1.80238109E-01     0.00018550   0.00024080   0.47673105   0.52217198   0.47625071  -0.36288291   0.11271919
  8.21043655E-02    -0.00008170  -0.00024139   0.16859589   0.21536222   0.25600883  -0.37686010  -0.58299714
  3.54288527E-02     0.00002271   0.00004912  -0.00189440   0.00026765   0.00307811  -0.01009148  -0.02810431
F
  1.07788524E+03     0.00066139  -0.00019180   0.00062963  -0.00123311   0.00168667   0.00252560
  3.65409080E+02     0.00611650  -0.00178165   0.00516386  -0.00914729   0.01105597   0.01321445
  1.55254800E+02     0.03221673  -0.00924938   0.02355697  -0.04020323   0.05116414   0.07446308
  7.31689300E+01     0.10757267  -0.03129252   0.08037328  -0.13224258   0.15072076   0.16488129
  3.66856800E+01     0.23886728  -0.06756767   0.15521256  -0.25005590   0.32161297   0.52195480
  1.92825954E+01     0.32998174  -0.08901212   0.22808388  -0.27860704   0.14563047  -0.39905538
  1.04209100E+01     0.32230710  -0.06439913   0.04886440   0.34723392  -0.70315325  -1.02578092
  5.54947315E+00     0.19639736   0.04495256  -0.27419029   0.52302944  -0.24474715   1.35136455
  2.85462934E+00     0.05633795   0.22284324  -0.58380604  -0.06922958   0.96931251  -0.01548639
  1.36162382E+00     0.00420808   0.36510976  -0.07657812  -0.69021627  -0.26060962  -1.17681923
  6.09628582E-01    -0.00000225   0.37908671   0.41344367   0.02083569  -0.74710139   1.16486834
  2.52838789E-01     0.00001167   0.25199346   0.29412640   0.50281350   0.62467306  -0.33844950
  9.13210536E-02    -0.00000244   0.06323981   0.07431200   0.15946321   0.26560814  -0.35504014
  5.04047100E-02     0.00000115  -0.00870342  -0.00969336  -0.01783575  -0.02862745   0.05272103
G
  5.48042940E+01    -0.05216008   0.10975667   0.11900634   0.23010759
  2.14723650E+01    -0.20852245   0.44089722   0.43535446   0.48807939
  8.73159940E+00    -0.25123808   0.36669364   0.11004711  -0.60666480
  3.97082170E+00    -0.31688226   0.03676084  -0.65337212  -0.41409547
  1.94839610E+00    -0.32915829  -0.24662569  -0.20547070   0.89571072
  8.09472840E-01    -0.28768056  -0.42806653   0.52049065  -0.12701185
  3.62200630E-01    -0.07065785  -0.12163762   0.24753720  -0.47578615
  1.25405870E-01    -0.01223940  -0.02001977   0.03267244  -0.07112408
\end{verbatim}
}

\newpage

{\small\begin{verbatim}
H
  3.86539440E+01     0.04576024  -0.20059738  -0.30440567
  1.48410360E+01     0.09884256  -0.47026800  -0.56340601
  7.05642780E+00     0.24622056  -0.25336193   0.20847715
  3.35510090E+00     0.38917461  -0.21780230   0.65127231
  1.71754750E+00     0.42826204   0.49893232  -0.31657951
  5.09950300E-01     0.26424815   0.41994691  -0.57505676
  2.77580970E-01    -0.06986946  -0.09564673   0.11552375
I
  9.60000000E+00     0.21662787
  4.80000000E+00     0.40293423
  2.40000000E+00     0.40351893
  1.20000000E+00     0.24229589
  6.00000000E-01     0.08603710
\end{verbatim}
}

\bibliography{bibfile}